\documentstyle[11pt,epsf]{article}

\newcommand{\newsection}{ \setcounter{equation}{0} \section}

\newcommand{\beq}{\begin{equation}} \newcommand{\eeq}{\end{equation}}
\newcommand{\bea}{\begin{eqnarray}} \newcommand{\eea}{\end{eqnarray}}

  \newcommand
{\Romannumeral}[1]{\uppercase\expandafter{\romannumeral#1}}

\newcommand{\be}{\begin{enumerate}} \newcommand{\ee}{\end{enumerate}}
\newcommand{\bi}{\begin{itemize}} \newcommand{\ei}{\end{itemize}}
\newcommand{\ba}{\begin{array}} \newcommand{\ea}{\end{array}}
\newcommand{\bc}{\begin{center}} \newcommand{\ec}{\end{center}}
\newcommand{\bt}{\begin{tabular}} \newcommand{\et}{\end{tabular}}

\def\lsim{\mathrel{\rlap{\lower4pt\hbox{\hskip1pt$\sim$}}
    \raise1pt\hbox{$<$}}}           
\def\gsim{\mathrel{\rlap{\lower4pt\hbox{\hskip1pt$\sim$}}
    \raise1pt\hbox{$>$}}}           

\newcommand{\tr}{\mathop{\rm tr}}           
\newcommand{\half}{\textstyle {1\over2} \displaystyle}    
\newcommand{\quarter}{\textstyle {1\over4} \displaystyle} 
\newcommand{\Dslash}{{\hbox{D}\kern-0.6em\raise0.15ex\hbox{/}}} 

\renewcommand{\et}{\eta}

\begin{document}

\setlength{\oddsidemargin}{0cm} \setlength{\baselineskip}{7mm}

\input epsf

\begin{normalsize}\begin{flushright}

September 2011 \\

\end{flushright}\end{normalsize}

\begin{center}
  
\vspace{5pt}

{\Large \bf Discrete Wheeler-DeWitt Equation}

\vspace{30pt}

{\sl Herbert W. Hamber}
$^{}$\footnote{e-mail address : HHamber@uci.edu} \\
Institut des Hautes Etudes Scientifiques \\
35, route de Chartres \\
91440 Bures-sur-Yvette, France. \\

\vspace{5pt}

and

\vspace{5pt}

{\sl Ruth M. Williams}
$^{}$\footnote{e-mail address : R.M.Williams@damtp.cam.ac.uk} \\
Department of Applied Mathematics and Theoretical Physics, \\
Wilberforce Road, Cambridge CB3 0WA, United Kingdom. \\

and 

Girton College, Cambridge CB3 0JG, United Kingdom. \\

\vspace{10pt}

\end{center}

\begin{center} {\bf ABSTRACT } \end{center}

\noindent

We present a discrete form of the Wheeler-DeWitt  equation for quantum gravitation,
based on the lattice formulation due to Regge.
In this setup the infinite-dimensional manifold of 3-geometries is replaced by
a space of three-dimensional piecewise linear spaces, 
with the solutions to the lattice equations providing a suitable approximation
to the continuum wave functional.
The equations incorporate a set of constraints on the quantum wavefunctional, arising
from the triangle inequalities and their higher dimensional analogs.
The character of the solutions is discussed in the strong coupling (large $G$) limit,
where it is shown that the wavefunctional only depends on geometric quantities, such
as areas and volumes.
An explicit form, determined from the discrete wave equation
supplemented by suitable regularity conditions, shows peaks corresponding to
integer multiples of a fundamental unit of volume.
An application of the variational method using correlated product wavefunctions
suggests a relationship between quantum gravity in $n+1$ dimensions, and 
averages computed in the Euclidean path integral formulation in $n$ dimensions.
The proposed discrete equations could provide a useful, and complementary,
computational alternative to the Euclidean lattice path integral approach to 
quantum gravity.





\vfill

\pagestyle{empty}

\newpage

\pagestyle{plain}

\section{Introduction}
\label{sec:intro}

\vskip 20pt

In this paper we will present a lattice version of the Wheeler-DeWitt equation of
quantum gravity. 
The approach used here will be rooted in the canonical formulation of quantum gravity, and
can therefore be regarded as complementary to the Euclidean lattice version of the same theory
discussed elsewhere.
In the following we will derive a discrete form of the Wheeler-DeWitt equation
for pure gravity,
based on the simplicial lattice formulation of gravity developed by Regge and Wheeler.
It is expected that the resulting lattice equation will reproduce the original continuum
equation in some suitable small lattice spacing limit.
In this formulation the infinite-dimensional manifold of 3-geometries is replaced by
the space of three-dimensional piecewise linear spaces, 
with solutions to the lattice equations then providing a suitable approximation
to the continuum gravitational wavefunctional.
The lattice equations will provide a new set of constraints on the quantum wavefunctional,
which arise because of the imposition of the triangle inequalities and their higher 
dimensional analogs.
The equations are explicit enough to allow a number of potentially useful practical calculations
in the quantum theory of gravity,
such as the strong coupling expansion, the weak field expansion,
mean field theory, and  the variational method.
In this work we will provide a number of sample calculations to 
illustrate the workings of the lattice theory, and what in our opinion is
the likely physical interpretation of the results.

In the strong coupling (large $G$) limit we will show that
the wavefunctional depends on geometric quantities only, such
as areas, volumes and curvatures, and that in this limit the correlation length is 
finite in units of the lattice spacing.
An explicit form of the wavefunctional, determined from the discrete equation
supplemented by suitable regularity conditions, shows peaks corresponding to
integer multiples of a fundamental unit of volume.
Later the variational method will be introduced, based here on correlated product 
(Jastrow-Slater type) wavefunctions.
This approach brings out a relationship between ground state properties of
quantum gravity in $n+1$ dimensions, and 
certain averages computed in the Euclidean path integral formulation in $n$ dimensions,
i.e. in one dimension {\it less}.
Because of its reliance on a different set of approximation methods, the 3+1 lattice
formulation presented here could provide a useful, and complementary, computational 
alternative to the Euclidean lattice path integral  approach to quantum gravity in four dimensions.
The equations are explicit enough that numerical solutions should be achievable in
a number of simple cases, such as a toroidal regular lattice with $N$ vertices in 3+1 dimensions.

An outline of the paper is as follows.
In Section 2, as a background to the rest of the paper, we describe the formalism of classical gravity, as set up by Arnowitt, Deser and Misner. In Section 3, we introduce the continuum form of the Wheeler-DeWitt equation and, in Section 4, describe how it can be solved in the minisuperspace approximation. Section 5 is the central core of the paper, where we transcribe the Wheeler-DeWitt equation to the lattice. Practical details of the lattice version are given in Section 6 and the equation solved in the strong coupling limit in both 2+1 and 3+1 dimensions. A general solution at the full range of couplings requires the inclusion of the curvature term, which was neglected in the strong coupling expansion, and Sections 7 and 8 outline methods of including this term, by perturbation theory and by the variational method. 
Section 9 gives a short outline of the lattice weak field expansion as it applies to the Wheeler-DeWitt
equation. Section 10 concludes with a discussion.

\vskip 40pt
\section{Arnowitt-Deser-Misner (ADM) Formalism and Hamiltonian}
\label{sec:adm}

Since this paper involves the canonical quantization of gravity \cite{dir58},
we begin with a discussion of the classical canonical formalism
derived by Arnowitt, Deser and Misner \cite{adm62}.
While many of the results presented in this section are rather well known, it will be
useful, in view of later applications, to recall the main results and formulas and
provide suitable references for expressions used later in the paper.

The first step in developing a canonical formulation for gravity is to introduce a time-slicing of
space-time, by introducing a sequence of spacelike hypersurfaces labeled by a continuous time coordinate $t$.
The invariant distance is then written as
\beq
d s^2 \; \equiv \; - d \tau^2 \, = \, g_{\mu\nu} \, dx^\mu dx^\nu 
 \, = \, g_{ij} \, d x^i  \, d x^j  \, + \, 2 g_{ij} \, N^i dx^j dt 
\, - \, ( N^2 \, - \, g_{ij} \, N^i N^j ) dt^2 \; ,
\eeq
where $x^i$ $(i=1,2,3)$ are coordinates on a three-dimensional manifold and $\tau$ is
the proper time, in units with $c=1$.
The relationship between the quantities 
$d \tau$, $dt$, $d x_i$, $N$ and $N_i$ basically
expresses the Lorentzian version of Pythagoras' theorem.

Indices are raised and lowered with $g_{ij} ( {\bf x} ) $ $(i,j=1,2,3)$, which denotes the three-metric on the
given spacelike hypersurface, and $N( {\bf x} )$ and $N^i ( {\bf x} )$ are the lapse and shift functions, respectively.
These last two quantities describe the lapse of proper time ($N$) between two infinitesimally close
hypersurfaces, and the corresponding shift in spatial coordinate ($N^i$). 
It is customary to mark four-dimensional quantities by the prefix ${}^4$, so that all un-marked quantities
will refer to three dimensions (and are occasionally marked 
explicitly by a ${}^3$).
In terms of the original four-dimensional metric ${}^4 g_{\mu\nu}$ one has
\beq
\left (  \matrix{
{}^4 g_{00} & {}^4 g_{0j} \cr
{}^4 g_{i0} & {}^4 g_{ij}  
}  \right )
\; = \; 
\left (  \matrix{
N_k N^k - N^2 & N_j  \cr
N_i  & g_{ij} 
}  \right )  \;\; ,
\eeq
and for its inverse
\beq
\left (  \matrix {
{}^4 g^{00} & {}^4 g^{0j} \cr
{}^4 g^{i0} & {}^4 g^{ij}  
}  \right )
\; = \; 
\left (  \matrix {
- 1 / N^2 & N^j / N^2  \cr
N_i / N^2  &  g^{ij} - N^i N^j /  N^2  
}  \right ) \;\; ,
\eeq
which then gives for the spatial metric and the lapse and shift functions
\beq
g_{ij} \; = \; {}^4 g_{ij} \;\;\;\;\;\;
N \; = \;  \left ( - {}^4 g^{00} \right )^{-1/2} \;\;\;\;\;\;
N_i \; = \; {}^4 g_{0i} \; .
\eeq
For the volume element one has
\beq
\sqrt{ \, {}^4 g } \; = \; N \, \sqrt{g} \; ,
\eeq
where the latter involves the determinant of the three-metric,
$g \equiv \det g_{ij}$.
As usual $g^{ij}$ denotes the inverse of the matrix $g_{ij}$.
In terms of these quantities, the Einstein-Hilbert Lagrangian of general relativity can then be written,
up to an overall multiplicative constant, in the following (first-order) form
\beq
{\cal{L}} \; = \; \sqrt{ {}^4 g } \;\;  {}^4 R \; = \; 
- \, g_{ij} \, \partial_t \pi^{ij}  \; - \;  N \, R^0 \; - \; N_i \, R^i  \; ,
\label{eq:adm-eh}
 \eeq
up to boundary terms.
Here one has defined the following quantities:
\bea
\pi^{ij} & \equiv & \sqrt{ {}^4 g } \, \left ( 
\, {}^4 \Gamma^0_{kl} - g_{kl} \, {}^4 \Gamma^0_{mn} \, g^{mn} 
\right ) g^{ik} g^{lj}
\nonumber \\
R^0 & \equiv& - \sqrt{g} \, \left [  
\, {}^3 \! R \, + \, g^{-1} ( \half \pi^2 - \pi^{ij} \pi_{ij} ) 
\right ]
\nonumber \\
R^i & \equiv & - 2 \, \nabla_j \, \pi^{ij} \; .
\label{eq:def-r0-ri}
\eea
The symbol $\nabla^i $ denotes covariant differentiation with respect
to the index $i$ using the spatial three-metric $g_{ij}$, and ${}^3 R$ is the scalar curvature computed
from this metric.
Also note that the affine connection coefficients $\Gamma^k_{ij}$ 
have been eliminated in favor of the spatial derivatives of the metric 
$\partial_k g_{ij}$, and one has defined $\pi = \pi^i_{\;\; i}$.
Since the quantities $N$ and $N^i$ do not appear in the 
$ \pi^{ij}  \, \partial_t g_{ij} $ part, they are interpreted as
Lagrange multipliers, and the "Hamiltonian" density
\beq
{\cal H } \; \equiv \; N \, R^0 \, + \, N_i R^i
\eeq
vanishes as a result of the constraints.
Varying the first order Lagrangian of Eq.~(\ref{eq:adm-eh}) with respect to $g_{ij}$, $N_i$, $N$ and
$\pi_{ij}$ one obtains a set of equations which are equivalent to Einstein's field equations.
First varying with respect to $\pi_{ij}$ one obtains an equation which 
can be viewed as defining $\pi_{ij}$,
\beq
\partial_t g_{ij} \; =\; 2 N g^{-1/2} \, ( \pi_{ij} - \half g_{ij} \, \pi ) 
\, + \, \nabla_j \, N_i \, + \, \nabla_i N_j \; .
\label{eq:adm-eq1}
\eeq
Varying with respect to the spatial metric $g_{ij}$ gives
the time evolution for $\pi_{ij}$ ,
\bea
\partial_t \pi^{ij} & = & 
- N \sqrt{g} \, ( \, {}^3 \! R^{ij} - \half g^{ij} \; {}^3 \! R )
+ \half N g^{-1/2} g^{ij} ( \pi^{kl} \pi_{kl} - \half \pi^2 ) 
\nonumber \\
&& - 2 N g^{-1/2} ( \pi^{ik} \pi_{k}^j - \half \pi \pi^{ij} ) 
+ \sqrt{g} \, ( \nabla^i \nabla^j N - g^{ij} \, \nabla^k \, \nabla_k \, N ) 
\nonumber \\
&& + \nabla_k \, ( \pi^{ij} N^k ) 
- \nabla_k \, N^i  \, \pi^{kj} - \nabla_k \, N^j \, \pi^{ki} \; .
\label{eq:adm-eq2}
\eea
Finally varying with respect to the lapse ($N$) and shift ($N^i$) functions gives
\beq
R^0 (g_{ij}, \pi_{ij}) =0 \;\;\;\;\;\;
R^i (g_{ij}, \pi_{ij}) =0 \; ,
\eeq
which can be viewed as the four constraint equations 
$ {}^4 G^0_\mu  = {}^4 \! R^0_\mu - \half \delta^0_\mu \; {}^4 \! R =0$, expressed for this choice
of metric decomposition \cite{dir58}.
The above constraints can therefore be considered as analogous to Gauss's law  
$\partial_i \, F^{i0} = {\bf \nabla \cdot E} =0$ in electromagnetism.

Some of the quantities introduced above (such as ${}^3 R$) describe intrinsic properties
of the spacelike hypersurface, while some others can be related to the extrinsic 
properties of such a hypersurface when embedded in
four-dimensional space.
If  spacetime is sliced up (foliated) by a  one-parameter 
family of spacelike hypersurfaces $x^\mu = x^\mu (x^i,t)$, then
one has for the intrinsic metric within the spacelike 
hypersurface
\beq
g_{ij} \; = \; g_{\mu\nu} \, X_i^\mu X_j^\nu  \;\;\;\;
{\rm with} \;\;\;\; X_i^\mu \; \equiv \; \partial_i \, x^\mu \; ,
\eeq
while the extrinsic curvature is given in terms of the unit normals
to the spacelike surface, $U^\mu$,
\beq
K_{ij} (x^k ,t) \; = \; - (\nabla_\mu \, U_{\nu}) \, X_i^\nu X_j^\mu   \; .
\eeq
In this language, the lapse and shift functions appear in the expression
\beq
\partial_t  x^\mu  \; = \; N \, U^\mu \, + \, N^i \, X^\mu_i \; .
\eeq
In the following $K= g^{ij} K_{ij} = K^i_{\;\; i}$ the trace of the matrix ${\bf K}$.

Now, in the canonical formalism, the momentum can be expressed
in terms of the extrinsic curvature
\beq
\pi^{ij} \; = \; - \, \sqrt{g}  \, ( K^{ij} - K \, g^{ij} ) \; .
\label{eq:mom-def}
\eeq
It is then convenient to define the quantities $H$ and $H_i$ as (here $\kappa = 8 \pi G$)
\bea
H & \equiv & 2 \, \kappa \, g^{-1/2} 
\left ( \pi_{ij} \pi^{ij} \, - \, \half \pi^2 \right ) \, - \, 
{ 1 \over 2 \, \kappa } \, \sqrt{g}   \; {}^3 \! R 
\nonumber \\
H_i  & \equiv &  - 2 \, \nabla_j  \, \pi_i^j  \; .
\label{eq:h-eq}
\eea
The last two statements are essentially equivalent to
the definitions in Eq.~(\ref{eq:def-r0-ri}).

In this notation, the Einstein field equations in the absence of sources
are equivalent to the initial value constraint
\beq
H(x) \; = \; H_i (x) \; = \; 0 \; ,
\label{eq:hamx}
\eeq
supplemented by the canonical evolution equations for  $\pi^{ij}$ and $g_{ij}$.
The quantity 
\beq
{\bf H } \; = \; \int d^3 x \; \left [  N (x) \, H(x) + N^i (x) H_i (x) \right ]
\eeq
should then be regarded as the Hamiltonian for classical 
general relativity.

When matter is added to the Einstein-Hilbert Lagrangian,
\beq
I [g,\phi] \; = \; {1 \over 16 \pi G } \,  \int d^4 x \; \sqrt{g} \;
{}^4 \! R (g_{\mu\nu} (x) ) \; + \; I_\phi [g_{\mu\nu} , \phi  ] \;\; ,
\eeq
where $\phi (x)$ are some matter fields,
the action within the ADM parametrization of the metric coordinates needs to be
modified to
\beq
I [g,\pi,\phi, \pi_\phi, N ] \; = \; 
\int dt \, d^3 x \left ( {\textstyle {1 \over 16 \pi G }  \displaystyle} 
\; \pi^{ij} \; \partial_t g_{ij} + \pi_\phi \, \partial_t \phi - N \, T - N^i T_i \right ) \; .
\eeq
One still has the same definitions as before for the (Lagrange multiplier) lapse and shift function,
namely 
$N= (- {}^4 \! g^{00} )^{-1/2} $ and $N^i = g^{ij} \; {}^4 \! g_{0j} $.
The gravitational constraints are modified as well, since now
one defines
\bea
T &  \equiv & {\textstyle {1 \over 16 \pi G } \displaystyle} 
 \, H ( g_{ij} , \pi^{ij} ) \; + \; 
H^\phi (g_{ij} , \pi^{ij}, \phi, \pi_\phi )
\nonumber \\
T_i & \equiv & {\textstyle {1 \over 16 \pi G } \displaystyle} 
 \, H_i ( g_{ij} , \pi^{ij} ) \; + \; 
H^\phi_i (g_{ij} , \pi^{ij}, \phi, \pi_\phi ) \; ,
\eea
with the first part describing the gravitational part  given earlier in 
Eq.~(\ref{eq:h-eq}),
\bea
 H ( g_{ij} , \pi^{ij} ) & = &  \, G_{ij,kl}  \; \pi^{ij} \pi^{kl} \, - \, 
\sqrt{g} \; {}^3 \! R \, + \, 2 \lambda \, \sqrt{g}
\nonumber \\
H_i ( g_{ij} , \pi^{ij} ) & = & - 2 \, g_{ij} \nabla_k  \, \pi^{jk} \; ,
\label{eq:ham}
\eea
here conveniently re-written using the (inverse of) 
the DeWitt supermetric
\beq
G_{ij,kl} \; = \;  \half \, g^{-1/2} \left ( 
g_{ik} g_{jl} + g_{il} g_{jk} + \alpha \, g_{ij} g_{kl} \right ) \;\; ,
\label{eq:dewitt-metric3d-inv}
\eeq
with parameter $\alpha = -1 $.
Note that in the previous expression a cosmological term
(proportional to $\lambda$) has been added as well,
for future reference.
For the matter part one has
\bea
 H^\phi ( g_{ij} , \pi^{ij} , \phi, \pi_\phi) & = &  
\sqrt{g} \; T_{00} \; (  g_{ij} , \pi^{ij}, \phi, \pi_\phi )
\nonumber \\
H^\phi_i ( g_{ij} , \pi^{ij} , \phi, \pi_\phi) & = & - \, \sqrt{g} \;
T_{0i} \; (  g_{ij} , \pi^{ij}, \phi, \pi_\phi ) \; .
\eea
We note here that the (inverse of the) DeWitt supermetric in 
Eq.~(\ref{eq:dewitt-metric3d-inv})
is also customarily used to define a distance in the space of three-metrics
$g_{ij} (x)$. 
Consider an infinitesimal displacement of such a metric
$g_{ij} \rightarrow g_{ij} + \delta g_{ij}$, and define
the natural metric $G$ on such deformations by considering a
distance in function space
\beq
\Vert \delta g \Vert^2 
\; = \; \int d^3 x \, N(x) \; G^{ij,kl} (x) \; \delta g_{ij} (x) \, \delta g_{kl} (x) \; .
\eeq
Here the lapse $N(x)$ is an essentially arbitrary but positive
function, to be taken equal to one in the following.
The quantity $ G^{ij,kl} (x) $ is the three-dimensional version
of the DeWitt supermetric,
\beq
G^{ij,kl} \; = \;  \half \, \sqrt{g} \, \left ( 
g^{ik} g^{jl} + g^{il} g^{jk} + \bar{\alpha} \, g^{ij} g^{kl} \right ) \;\; ,
\label{eq:dewitt-metric3d}
\eeq
with the parameter $\alpha$ of Eq.~(\ref{eq:dewitt-metric3d-inv})
related to $\bar{\alpha}$ in Eq.~(\ref{eq:dewitt-metric3d}) by 
$\bar{\alpha} = - 2 \alpha / ( 2 + 3 \alpha ) $, so that
$\alpha = -1 $ gives $\bar{\alpha} = -2$ (note that this is
dimension-dependent).

\vskip 40pt
\section{Wheeler-DeWitt Equation}
\label{sec:wdweq}

Within the framework of the previous construction, a transition
from a classical to a quantum description of gravity is obtained
by promoting $g_{ij}$, $\pi^{ij}$, $H$ and $H_i$ to quantum operators, with ${\hat g}_{ij}$ and ${\hat
\pi}^{ij}$ satisfying canonical commutation relations.
In particular the classical constraints now select a physical vacuum
state $ \vert \Psi \rangle $, such that in the source free case
\beq
{\hat H} \, \vert \Psi \rangle \; = \; 0
\;\;\;\;\;\;
{\hat H}_i \, \vert \Psi \rangle \; = \; 0 \;\; ,
\eeq
and in the presence of sources more generally
\beq
{\hat T} \, \vert \Psi \rangle \; = \; 0
\;\;\;\;\;\;
{\hat T}_i \, \vert \Psi \rangle \; = \; 0 \;\; .
\label{eq:quant-const}
\eeq
As in ordinary nonrelativistic quantum mechanics, one can
choose different representations for the canonically conjugate
operators ${\hat g}_{ij}$ and ${\hat \pi}^{ij}$.
In the functional {\it position representation} one sets
\beq
{\hat g}_{ij} ( {\bf x} ) \; \rightarrow \;   g_{ij} ( {\bf x} ) 
\;\;\;\;\;\;
{\hat \pi}^{ij} ( {\bf x} ) \; \rightarrow \;  
- i \hbar \cdot16 \pi G \cdot
{ \delta \over \delta g_{ij} ( {\bf x} )  } \; .
\eeq
In this picture the quantum states become wave functionals of the 
three-metric $ g_{ij} ({\bf x} ) $,
\beq
\vert \Psi \rangle \; \rightarrow \;  \Psi \, [ g_{ij} ( {\bf x} ) ] \; .
\eeq
The two quantum constraint equations in Eq.~(\ref{eq:quant-const})
then become the Wheeler-DeWitt equation \cite{dew64,dew67,whe68}
\beq
\left \{ -  \, 16 \pi G \cdot G_{ij,kl}  \, 
{ \delta^2 \over \delta g_{ij} \, \delta g_{kl}  } \; - \;
{ 1 \over 16 \pi G } \, \sqrt{g} \left ( \,  {}^3 \! R \, - \, 2 \lambda \, \right ) \, + \,  {\hat
H}^\phi
\right \} \; 
\Psi [ g_{ij} ( {\bf x} ) ] \; = \; 0 \; ,
\label{eq:wd-1}
\eeq
with the inverse supermetric given in 3+1 dimensions by
\beq
G_{ij,kl} \; = \; 
\half \, g^{-1/2} \, \left (
g_{ik} g_{jl} + g_{il}  g_{jk} - g_{ij} g_{kl} \right ) \; ,
\eeq
and the diffeomorphism (or momentum) constraint
\beq
\left \{ 2 \, i  \, g_{ij}  \, \nabla_k \,
 { \delta \over \delta g_{jk}  } \, + \,
{\hat H}^\phi_i 
\right \} \; 
\Psi [ g_{ij} ( {\bf x} )  ] 
\; = \; 0 \; .
\label{eq:wd-2}
\eeq
This last constraint implies that the gradient of $\Psi$ on the 
superspace of $g_{ij}$'s and $\phi$'s is zero along those directions
that correspond to gauge transformations, i.e. diffeomorphisms 
on the three dimensional manifold, whose points
are labeled by the coordinates ${\bf x}$.
The lack of covariance of the ADM approach has not gone away,
and is therefore still part of the present formalism.
Also note that the DeWitt supermetric is not positive definite, which implies
that some derivatives with respect to the metric have the ''wrong'' sign.
It is understood that these directions correspond to the conformal mode.

A number of basic issues need to be addressed before one
can gain a full and consistent understanding of the dynamical
content of the theory \cite{leu64,kuc76,kuc92,ish93,bar98}.
These include possible problems of operator ordering, and
the specification of a suitable Hilbert space, 
which entails at some point a choice for the inner product 
of wave functionals,  for example in the Schr\"odinger form
\beq
\langle \Psi \vert \Phi \rangle \; = \; 
\int  d \mu [g] \; \Psi^{*} [ g_{ij} ] \, \Phi [ g_{ij} ] \,
\eeq
where $ d \mu [g] $ is some appropriate measure over
the three-metric $g$.
Note also that the continuum Wheeler-DeWitt equation contains, in the
kinetic term, products of functional differential operators which are evaluated 
at the same spatial point ${\bf x}$.
One would expect that such terms could produce $\delta^{(3)} (0)$-type singularities
when acting on the wave functional, which would then have to be regularized in some way.
The lattice cutoff discussed below is one way to provide such an explicit regularization.

A peculiar property of the Wheeler-DeWitt equation,
which distinguishes it from the usual Schr\"odinger equation
$H \Psi = i \hbar \partial_t \Psi $, 
is the absence of an explicit time coordinate.
As a result the r.h.s. term of the Schr\"odinger equation is
here entirely absent.
The reason is of course diffeomorphism invariance of the
underlying theory, which
expresses now the fundamental quantum equations in terms of 
fields $g_{ij}$, and not coordinates.
Consequently the Wheeler-DeWitt equation contains no
explicit time evolution parameter.
Nevertheless in some cases it seems possible to
assign the interpretation of "time coordinate"
to some specific variable entering the Wheeler-DeWitt
equation, such as the overall spatial volume or the
magnitude of some scalar field \cite{ish93}.

We shall not discuss here the connection between the Wheeler-DeWitt
equation and the Feynman path integral for gravity.
In principle any solution of the Wheeler-DeWitt equation 
corresponds to a possible quantum state of the universe.
A similar situation already arises, of course in much simpler form,
 in nonrelativistic quantum mechanics \cite{fey65}.
The effects of the boundary conditions on the
wavefunction will then act to restrict the class of possible 
solutions; in ordinary quantum mechanics these are determined by the 
physical context of the problem and some set of external
conditions.
In the case of the universe the situation is far less clear, 
and in many
approaches some suitable set of boundary conditions need to be postulated,
based on general arguments involving simplicity or economy.
One proposal \cite{har83} is to restrict the
quantum state of the universe by requiring that the wave function
$\Psi$ be determined by a path integral over compact Euclidean metrics.
The wave function would then be given by
\beq
\Psi [ g_{ij}, \phi ] \; = \; 
\int [ d g_{\mu\nu} ] \, [ d \phi ] \;
\exp \left ( - \hat{I} [  g_{\mu\nu} , \phi ] \right ) \;\; ,
\eeq
where $\hat{I}$ is the Euclidean action for gravity plus matter
\beq
\hat{I} \; = \; - { 1\over 16 \pi G } \int d^4 x \sqrt{g} \, ( R - 2 \lambda )
\, - \, { 1 \over 8 \pi G } \int d^3 x \sqrt{g_{ij} } \, K
\, - \, \int d^4 x \sqrt{g} \, {\cal L}_m \; .
\eeq
The semi-classical functional integral would then be restricted to those four-metrics
which have the induced metric $g_{ij}$ and the matter field $\phi$
as given on the boundary surface $S$.
One would then expect (as in the case in nonrelativistic quantum mechanics, where the path integral
with a boundary surface satisfies the Schr\"odinger equation),  that the
wavefunction constructed in this way would also automatically satisfy the Wheeler-DeWitt equation,
and this is indeed the case.

\vskip 40pt
\section{Minisuperspace}
\label{sec:minisup}

Minisuperspace models can in part provide an additional motivation for
our later work.
The quantum  state of a gravitational system is described, in
the Wheeler-DeWitt framework just introduced, by a wave function $\Psi$ 
which is a functional of the three-metric $g_{ij}$ and the matter
fields $\phi$. 
In general the latter could contain fields of arbitrary spins, but
here we will consider for simplicity just one single 
component scalar field $\phi (x)$.
The wavefunction $\Psi$  will then obey
the zero energy Schr\"odinger-like equation of Eqs.~(\ref{eq:wd-1})
and (\ref{eq:wd-2}).
The quantum state described by $\Psi$ is then a functional on the infinite-dimensional 
manifold $W$ consisting of all positive definite metrics $g_{ij} (x)$ and
matter fields $\phi (x)$ on a spacelike three-surface $S$.
We note here that on this space there is a natural metric
\beq
ds^2 [ \delta g, \delta \phi ] \, = \, \int { d^3 x \, d^3 x' \over N(x) } \left [ 
G^{ij,kl} (x,x') \; \delta g_{ij} \, \delta g_{kl} (x') \, + \,
\sqrt{g} \, \delta^3 (x-x') \; \delta \phi (x) \delta \phi (x') \right ] \; ,
\label{eq:sup-met-wd}
\eeq
where
\beq
G^{ij,kl} (x,x') \; = \;  G^{ij,kl} (x) \, \delta^3 (x - x')
\eeq
and
\beq
G^{ij,kl} (x) \; = \; \half \, \sqrt{g} \left [
g^{ik} (x) g^{jl} (x) + g^{il} (x) g^{jk} (x) - 2 \, g^{ij} (x) g^{kl} (x) 
\right ]
\eeq
is the DeWitt supermetric.

In general the wavefunction for all the dynamical variables of the
gravitational field in the universe is difficult to calculate, since an
infinite number of degrees of freedom are involved: the infinitely
many values of the metric at all spacetime points, and the infinitely many values of the matter field
$\phi$ at the same points.
One option is to restrict the choice of variable to a finite number
of suitable degrees of freedom \cite{bly75,haw84,haw85,hap85,pag02}.
As a result the overall quantum fluctuations are severely restricted, 
since these are now only allowed to be nonzero along the surviving dynamical directions.
If the truncation is severe enough, the transverse-traceless nature
of the graviton fluctuation is lost as well.
Also, since one is not expanding the quantum solution in a small
parameter, it can be difficult to estimate corrections.

In a cosmological  context, it seems natural to consider initially a homogeneous and isotropic model, and
restrict the
function space to two variables, the scale factor $a(t)$ and  a
minimally coupled homogeneous scalar field $\phi(t)$ \cite{hap85}.
The space-time metric is given by
\beq
d \tau^2 \; = \;  N^2 (t) \, dt^2 \; - \;  g_{ij} \, dx^i \, dx^j  \;\; .
\eeq
The three-metric $g_{ij}$ is then determined entirely by
the scale factor $a(t)$, 
\beq
g_{ij} \; = \; a^2 (t) \; {\tilde g}_{ij}
\eeq
with ${\tilde g}_{ij}$ a time-independent 
reference three-metric with constant curvature, 
\beq
{}^3 \! {\tilde R}_{ijkl} \; = \; k \left ( 
{\tilde g}_{ij} \, {\tilde g}_{kl} \, - \, {\tilde g}_{il}  \, {\tilde g}_{jk}
\right ) \; ,
\eeq
and $k=0,\pm 1$ corresponding to the flat, closed and open
universe case respectively.
In this case the minisuperspace $W$ is two-dimensional, with
coordinates $a$ and $\phi$, and supermetric
\beq
ds^2 [ a, \phi ] \; = \; N^{-1} ( - a \, da^2 \, + \, a^3 \, d \phi^2 ) \; .
\eeq
From the above expression for $ds^2 [ a, \phi ] $ 
one obtains the Laplacian in the above metric, required for
the kinetic term in the Wheeler-DeWitt equation,
\footnote{
The ambiguity regarding the operator ordering of 
$p^2 /  a = a^{-(q+1)} p a^q p $ in the Wheeler-DeWitt
equation can in principle be retained by writing for the above operator $\nabla^2$ the expression
$ - ( N / a^{q+1} ) \left \{ 
( \partial / \partial a ) \, a^q \, ( \partial / \partial a )
\,  - \, ( \partial^2 / \partial \phi^2 ) 
\right \} $, but this does not seem to affect the qualitative 
nature of the solutions. The case discussed in the text
corresponds to $q=1$, but $q=0$ seems even simpler.
}
\beq
- \half \, \nabla^2 (a, \phi) \; = \; { N \over 2 \, a^2 } \left \{ 
{ \partial \over \partial a } \, a \, { \partial \over \partial a }
\,  - \, { 1 \over a } \, { \partial^2 \over \partial \phi^2 } \right \} \;\; .
\eeq
Since the space is homogeneous, the diffeomorphism constraint is trivially satisfied.
Also $N$ is independent of $g_{ij}$ so in the homogeneous case
it can be taken to be a constant, conveniently chosen 
as $N=1$. 

It should be clear that in general the quantum behavior of the 
solutions is expected to be quite different from
the classical one. In the latter case one imposes some
initial conditions on the scale factor at some time $t_0$, which
then determines $a(t)$ at all later times.
In the minisuperspace view of quantum cosmology one has
to instead impose a condition on the wavepacket $\Psi$ at
$a=0$.
Due to their simplicity, in general it is possible to analyze the 
solutions to the minisuperspace Wheeler-DeWitt equation 
in a rather complete way, given some sensible
assumptions on how $\Psi (a,\phi)$ should behave,
for example, when the scale factor $a$ approaches zero.

In concluding the discussion on minisuperspace models as a tool
for studying the physical content of the Wheeler-DeWitt equation
it seems legitimate though to ask the following question: to what extent can
results for these very simple models which involve such a drastic truncation of 
physical degrees of freedom,  be ultimately representative of, and physically 
relevant to, what might, or might not, happen in the full quantum theory?


\vskip 40pt
\section{Lattice Hamiltonian for Quantum Gravity}
\label{sec:ham-grav-latt}

In constructing a discrete Hamiltonian for gravity one has
to decide first what degrees of freedom one should
retain on the lattice.
There are a number of possibilities, depending on which
continuum theory one chooses to discretize, and at what stage.
So, for example, one could start with a discretized version of
Cartan's formulation, and define vierbeins and spin connections
on a flat hypercubic lattice. 
Later one could define the transfer matrix for such a theory, and
construct a suitable Hamiltonian.

Another possibility, which is the one we choose to pursue here,
is to use the more economical (and geometric) Regge-Wheeler lattice discretization 
for gravity \cite{reg61,whe64},
with edge lengths suitably defined on a random lattice as the primary 
dynamical variables.
Even in this specific case several avenues for discretization
are possible.
One could discretize the theory from the very beginning, while it
is still formulated in terms of an action, 
and introduce for it a lapse and a shift function, extrinsic and intrinsic discrete curvatures etc.
Alternatively one could try to discretize the continuum
Wheeler-DeWitt equation
directly, a procedure that makes sense in the lattice formulation,
as these equations are still given in terms of geometric objects,
for which the Regge theory is very well suited.
It is the latter approach which we will proceed to outline here.

The starting point for the following discussion
is therefore the Wheeler-DeWitt equation for pure gravity
in the absence of matter, Eq.~(\ref{eq:wd-1}), 
\beq
\left \{ \, -  \, (16 \pi G )^2 \, G_{ij,kl}  ( {\bf x} )  \, 
{ \delta^2 \over \delta g_{ij} ( {\bf x} ) \, \delta g_{kl}  ( {\bf x} ) } \; - \;
\sqrt{g ( {\bf x} ) } \; \left ( \;  
{}^3 \! R ( {\bf x} ) \, - \, 2 \lambda \, \right ) \, \right \} \; 
\Psi [ g_{ij} ( {\bf x} ) ] \; = \; 0
\label{eq:wd-1b}
\eeq
and the diffeomorphism constraint of Eq.~(\ref{eq:wd-2}),
\beq
\left \{ \, 2 \, i  \, g_{ij}  ( {\bf x} ) \, \nabla_k ( {\bf x} ) \,
 { \delta \over \delta g_{jk}  ( {\bf x} ) }  \, \right \} \; 
\Psi [ g_{ij} ( {\bf x} )  ] 
\; = \; 0 \; .
\label{eq:wd-2b}
\eeq
Note that these equations express a constraint on the state $ \vert \Psi \rangle$
at {\it every} ${\bf x}$, each of the form $ {\hat H} ( {\bf x} ) \, \vert \Psi \rangle = 0$ and
$ {\hat H}_i \, ( {\bf x} ) \vert \Psi \rangle = 0 $.

On a simplicial lattice \cite{rowi81,che82,itz83,hw84,har85} (see for example \cite{hbook}, and references therein,
for a more complete discussion of the lattice formulation for gravity)
one knows that deformations of the squared edge
lengths are linearly related to deformations of the induced
metric.
In a given simplex $\sigma$, take coordinates based at a vertex $0$, 
with axes along the edges from $0$.
The other vertices are each at unit coordinate distance from $0$ (see Figures 1,2 and 3 for this labelling of a triangle and of a tetrahedron). 
In terms of these coordinates, the metric within the simplex is given by
\beq
g_{ij} (\sigma) \; = \; \half \, 
\left ( l_{0i}^2 + l_{0j}^2 - l_{ij}^2 \right ) \;\; .
\label{eq:latmet-1}
\eeq
Note also that in the following discussion only edges and volumes
along the spatial direction are involved.
It follows that one can introduce in a natural way a lattice analog
of the DeWitt supermetric of Eq.~(\ref{eq:dewitt-metric3d}),
by adhering to the following procedure.
First one writes for the supermetric in edge length space 
\beq
\Vert \, \delta l^2 \, \Vert^2 \; = \; \sum_{ij} \; G^{ij} (l^2)
\; \delta l^2_i \; \delta l^2_j \; \; ,
\eeq
with the quantity $G^{ij} (l^2)$ suitably 
defined on the space of squared edge lengths \cite{lun74,har97}. 
Through a straightforward exercise
of varying the squared volume of a given simplex $\sigma$ in $d$
dimensions
\beq
V^2 ( \sigma ) \; = \; {\textstyle \left ( { 1 \over d! } \right )^2 \displaystyle}
\det g_{ij}(l^2( \sigma )) 
\eeq
to quadratic order in the metric (on the r.h.s.), or in the squared edge
lengths belonging to that simplex (on the l.h.s.), one finds the identity 
\beq
{1 \over V (l^2)} \; 
\sum_{ij} { \partial^2 V^2 (l^2) \over \partial l^2_i \partial l^2_j }
\; \delta l^2_i \; \delta l^2_j \; = \;
{\textstyle { 1 \over d! } \displaystyle} \sqrt{ \det ( g_{ij} ) }
\left [ \; g^{ij} g^{kl} \delta g_{ij} \delta g_{kl}
- g^{ij} g^{kl} \delta g_{jk} \delta g_{li} \; \right ] \;\; .
\eeq
The right hand side of this equation contains precisely
the expression appearing in the continuum supermetric
of Eq.~(\ref{eq:dewitt-metric3d})
(for a specific choice of the parameter $\bar{ \alpha} = -2$),
while the left hand side contains the sought-for lattice supermetric.
One is therefore led to the identification
\beq
G^{ij} (l^2) \; = \; - \; d! \; \sum_{\sigma} \;
{1 \over V (\sigma)} \; 
{ \partial^2 \; V^2 (\sigma) \over \partial l^2_i \; \partial l^2_j } \;\; .
\label{eq:inverselasup}
\eeq
It should be noted that in spite of the appearance of a sum over simplices $\sigma$, 
$G^{ij} (l^2) $ is quite local (in correspondence with 
the continuum, where it
is ultra-local), since the derivatives on the right hand side
vanish when the squared edge lengths in question are
not part of the same simplex.
The sum over $\sigma$ therefore only extends over those few 
tetrahedra which contain either the $i$ or the $j$ edge.

At this point one is finally ready to write a lattice analog of the
Wheeler-DeWitt equation for pure gravity, which reads 
\beq
\left \{ \, -  \, (16 \pi G)^2 \, G_{ij} ( l^2 ) \, 
{ \partial^2 \over \partial l^2_{i} \, \partial l^2_{j}  }
\; - \;
\sqrt{g (l^2) } \; \left [ \; 
{}^3 \! R (l^2) \, - \, 2 \lambda \; \right ] \; \right \} \; 
\Psi [ \, l^2 \, ] \; = \; 0 \;\; ,
\label{eq:wd-latt}
\eeq
with $G_{ij} (l^2)$ the inverse of the matrix $G^{ij} (l^2)$ given above.
The range of the summation over $i$ and $j$ and the appropriate expression
for the scalar curvature, in this equation, are discussed below and
made explicit in Eq.~(\ref{eq:wd-latt1}).

It should be emphasized that, just like there is one local equation for {\it each} spatial
point $\bf{x}$ in the continuum, here too there is only one local (or semi-local, since strictly 
speaking more than one lattice vertex is involved) equation that needs to be
specified at {\it each} simplex, or simplices, with $G_{ij}$ defined in accordance
with the definition in Eq.~(\ref{eq:inverselasup}).
On the other hand, and again in close analogy with the continuum expression,
the wavefunction $\Psi [ \, l^2 \, ]$ depends of course collectively on {\it all} the edge lengths 
in the lattice.
The latter should therefore be regarded as a function of the whole simplicial geometry,
whatever its nature might be,  just like the continuum wavefunction 
$ \Psi [ g_{ij} ] $ will be a function(al) of {\it all} metric variables,
or more specifically of the overall geometry of the manifold, due
to the built-in diffemorphism invariance.
On the side we note here that the lattice supermetric is dimensionful,
$G_{ij} \sim l^{4-d}$ and $G^{ij} \sim l^{d-4}$ in
$d$ spacetime dimensions, so
it might be useful and convenient from now on to explicitly introduce
a lattice spacing $a$ (or a momentum 
cutoff $\Lambda =1/a$) and express all dimensionful
quantities ($G,\lambda, l_i $) in terms of this 
fundamental lattice spacing.

As noted, Eqs.~(\ref{eq:wd-1}) or (\ref{eq:wd-latt})
both express a constraint equation at each \lq\lq point" in space.
Here we will elaborate a bit more on this point.
On the lattice, points in space are replaced by a set of edge labels $i$, 
with a few  edges clustered around each vertex, in a 
way that depends on the dimensionality and
the local lattice coordination number.
To be more specific, the first term in Eq.~(\ref{eq:wd-latt})
contains derivatives with respect to edges $i$ and $j$ 
connected by a matrix element $ G_{ij} $ which
is nonzero only if $i$ and $j$
are close to each other, essentially nearest neighbor.
One would therefore expect that the first term could 
be represented by just a sum of edge contributions,
all from within one ($d-1$)-simplex $\sigma$ (a tetrahedron in
three dimensions).
The second  term containing ${}^3 \! R (l^2)$ 
in Eq.~(\ref{eq:wd-latt}) is also local in the edge lengths:
it only involves a handful of edge lengths which enter
into the definition of areas, volumes and angles around
the point ${\bf x}$, and 
follows from the fact that the local curvature at the 
original point ${\bf x}$ is completely determined by the
values of the edge lengths clustered around $i$ and $j$.
Apart from some geometric factors, it describes,
through a deficit angle $\delta_h$, the parallel
transport of a vector around an elementary dual
lattice loop.
It should therefore be adequate
to represent this second term by a sum over contributions
over all ($d-3$)-dimensional hinges (edges in 3+1 dimensions) 
$h$ attached to the simplex $\sigma$,
giving therefore in three dimensions 
\beq
\left \{ \, -  \, (16 \pi G )^2 \sum_{ i,j \subset \sigma}
\, G_{ij} \, ( \sigma ) \, 
{ \partial^2 \over \partial l^2_{i} \, \partial l^2_{j}  }
\; - \; 2 \, n_{\sigma h} \; \sum_{ h \subset \sigma} \, l_h \, \delta_h 
\; + \; 2 \lambda  \; V_\sigma  \,
\right \} \; 
\Psi [ \, l^2 \, ] \; = \; 0 \;\; .
\label{eq:wd-latt1}
\eeq
Here $\delta_h$ is the deficit angle at the hinge $h$,
$l_h$ the corresponding edge length, $V_\sigma = \sqrt{ g(\sigma)} $
the volume of the simplex (tetrahedron in three spatial dimensions) 
labeled by $\sigma$.
$ G_{ij} \, ( \sigma ) $ is obtained either from Eq.~(\ref{eq:inverselasup}),
 or from the lattice transcription of Eq.~(\ref{eq:dewitt-metric3d-inv})
\beq
G_{ij,kl} (\sigma) \; = \;  \half \, g^{-1/2} (\sigma) \left [ 
g_{ik} (\sigma) g_{jl} (\sigma) + g_{il} (\sigma) g_{jk} (\sigma) - 
\, g_{ij} (\sigma) g_{kl} (\sigma) 
\right ] \; \; ,
\label{eq:dewitt-metric3d-inv-latt}
\eeq
with the induced metric $g_{ij} (\sigma) $ within a simplex $\sigma$
given in Eq.~(\ref{eq:latmet-1}).
The combinatorial factor $ n_{\sigma h} $ ensures the correct normalization
for the curvature term, since the latter has to give the lattice version of
$ \int \sqrt{g} \, {}^3 R = 2 \sum_h \delta_h l_h $  (in three spatial dimensions)
when summed over all simplices $\sigma$.
The inverse of $ n_{\sigma h} $ counts therefore the number of times
the same hinge appears in various neighboring simplices, and consequently depends 
on the specific choice of underlying lattice structure; 
for a flat lattice of equilateral triangles in two dimensions $ n_{\sigma h} =1/6 $ .
\footnote{
Instead of the combinatorial factor $n_{\sigma h} $ one could insert
a ratio of volumes $V_{\sigma h}/ V_h$(where $V_h$ is the volume per hinge \cite{hw84} and $V_{\sigma h}$ is the amount of that volume in the simplex $\sigma$),
but the above form is simpler.}
The lattice Wheeler-DeWitt equation given in Eq.~(\ref{eq:wd-latt1}) is the main
result of this paper.

It is in fact quite encouraging that the discrete equation in 
Eqs.~(\ref{eq:wd-latt}) and (\ref{eq:wd-latt1})  is very similar
to what one would derive in Regge lattice gravity by doing
the 3+1 split of the lattice metric carefully from the very
beginning  \cite{wil86,pir86,tuc90}.
These authors also derived a lattice Hamiltonian
in three dimensions, written in terms of lattice
momenta conjugate to the edge length variables.
In this formulation the Hamiltonian constraint 
equations have the form
\bea
H_n & = & \quarter \sum_{ \alpha \in n } \, 
G_{ij}^{ ( \alpha ) } \, \pi^i \, \pi^j \, - \, 
\sum_{ \beta \in n } \sqrt{ g_\beta } \; \delta_\beta
\nonumber \\
& = & \quarter 
\sum_{ \alpha \in n } \, { 1 \over V_\alpha } 
\, \left [ ( \tr  \, \pi^2 )_\alpha  - \half ( \tr \, \pi )_\alpha^2 \right ]
\, - \,  \sum_{ \beta \in n } \sqrt{ g_\beta } \; \delta_\beta 
\; = \; 0 \; ,
\eea
with $H_n$ defined on the lattice site $n$.
The sum  $\sum_{ \alpha \in n} $ extends over 
neighboring tetrahedra
labelled by $\alpha$, whereas the sum $ \sum_{ \beta \in n } $
extends over neighboring edges, here labeled by $\beta$.
$G_{ij}^{ ( \alpha ) } $ is the inverse of the DeWitt supermetric
at the site $\alpha$, and $\delta_\beta $ the deficit angle
around the edge $\beta$.
$\sqrt{ g_\beta }$ is the dual (Voronoi) volume associated
with the edge $\beta$.

The lattice Wheeler-DeWitt equation of Eq.~(\ref{eq:wd-latt})
has an interesting structure, which is in part reminiscent of the 
Hamiltonian for lattice gauge theories.
The first, local kinetic term is the gravitational analog of the 
electric field term $E_a^2 $.
It contains momenta which can be considered as conjugate
to the squared edge length variables.
The second local term involving ${}^3 \! R (l^2)$ is the analog
of the magnetic $ (\nabla \times A_a )^2$.
In the absence of a cosmological constant term, the first
and second term have opposite sign, and need to
cancel out exactly on physical states in order to give 
$ H ( {\bf x} ) \, \Psi =0 $.
On the other hand, the last term proportional to $\lambda$
has no gauge theory analogy, and is, as expected, 
genuinely gravitational.

It  seems important to note here that the squared edge lengths take
on only positive values 
$l_i^2 > 0$, a fact that would seem to imply
that the wavefunction has to vanish when the edge
lengths do, $\Psi ( l^2=0 ) \simeq 0$.
This constraint will tend to select the regular solution close to
the origin in edge length space, as will be discussed further below.
In addition one has some rather complicated further constraints 
on the squared edge lengths, due to the triangle 
inequalities. 
These ensure that the areas of triangles
and the volumes of tetrahedra are always positive.
As a result one would expect an average soft 
local upper bound on the
squared edge lengths of the type $l_i^2 \lsim l_0^2$
where $l_0$ is an average edge length,
$\langle l_i^2 \rangle = l^2_0$.
The term "soft" refers to the fact that while large values for
the edge lengths are possible, these should nevertheless
enter with a relatively small probability, due to the small phase
space available in this region.
In any case, the nature of the discrete Wheeler-DeWitt equation presented here
is explicit enough so that these, and other related, issues can presumably be 
answered both satisfactorily and unambiguously.

The above considerations have some consequences already
in the strong coupling limit of the theory.
For sufficiently strong coupling
(large Newton constant $G$) the first term in Eq.~(\ref{eq:wd-latt})
is dominant, which shows again some similarity with what one
finds for non-Abelian gauge theories for large coupling $g^2$.
One would then expect both from the constraint  $l_i >0 $ 
and the triangle inequalities, that the spectrum of this operator
is discrete, and that the energy gap, the spacing between the lowest
eigenvalue and the first excited state, is of the same order as the 
ultraviolet cutoff.
Nevertheless one important difference here is that one is not interested
in the whole spectrum, but instead just in the zero mode.

\begin{table}

\begin{center}
\begin{tabular}{|l|l|}
\hline\hline
&
\\
dimension & dimension of Laplacian $\Delta_g$
\\ \hline \hline
$d$ dimensions & $l^{4-d}/l^4 \sim l^{-d} \sim 1/ V_d $
\\ \hline
2+1 dimensions & $A/l^4 \sim 1/A $
\\ \hline
3+1 dimensions & $l/l^4 \sim 1/l^3 \sim 1/V $
\\ \hline \hline
\end{tabular}
\end{center}
\label{lap}

{\small {\it
Table I: Dimension of the Laplacian term in $d$ dimensions.
\medskip}}


\end{table}
\vskip 10pt

\begin{table}

\begin{center}
\begin{tabular}{|l|l|l|l|l|}
\hline\hline
& & & &
\\
dimension & $G$ dimension & $\lambda$ dimension & dimensional & dimensionless  
\\ \hline \hline
$d$ dimensions & $l^{d-2}$ & $1/l^2$ & $G / \sqrt{\lambda} \sim l^{d-1}$ & $G^{2/(d-2)} \lambda $
\\ \hline
2+1 dimensions & $l$ & $1/l^2$ & $G / \sqrt{\lambda} \sim A_\Delta $ & $G^2 \lambda $
\\ \hline
3+1 dimensions & $l^2$ & $1/l^2$ & $G / \sqrt{\lambda} \sim V_T $ & $G \lambda$
\\ \hline \hline
\end{tabular}
\end{center}
\label{dim}

{\small {\it
Table II: Dimensions of couplings in $d$ dimensions.
\medskip}}


\end{table}
\vskip 10pt

Irrespective of its specific form,
it is in general possible to simplify the lattice Hamiltonian
constraint in 
Eqs.~(\ref{eq:wd-latt}) and (\ref{eq:wd-latt1})
by using 
scaling arguments, as one does often in ordinary
nonrelativistic quantum mechanics (for a list of relevant dimensions see Table I
and Table II).
After setting for the scaled cosmological constant
$\lambda = 8 \pi G \, \lambda_0$ and dividing the
equation out by common factors, it
can be recast in the slightly simpler form
\beq
\left \{ \, -  \, \alpha \, a^6 \cdot
{ 1 \over \sqrt{ g (l^2) } } \; G_{ij} ( l^2 ) \, 
{ \partial^2 \over \partial l^2_{i} \, \partial l^2_{j}  }
\, - \, \beta \, a^2 \cdot {}^3 \! R (l^2) \, + \, 1 \, 
\right \} \; 
\Psi [ \, l^2 \, ] \; = \; 0 \;\; ,
\label{eq:wd-latt2}
\eeq
where one finds it useful to
define a dimensionless Newton's constant, as
measured in units of the cutoff $\bar{G} \equiv 16 \pi G /a^2 $, 
and a dimensionless cosmological constant
$\bar{\lambda_0} \equiv \lambda_0 a^4$, so that in
the above equation one has
$\alpha = \bar{G}  / \bar{\lambda}_0$
and $\beta = 1/ \bar{G}  \bar{\lambda_0}  $.
Furthermore the edge lengths have been rescaled
so as to be able to set $\lambda_0=1$ in lattice units
(it is clear from the original
gravitational action that the cosmological constant
$\lambda_0$ simply multiplies the total spacetime
volume, thereby just shifting around the overall scale
for the problem). 
Schematically Eq.~(\ref{eq:wd-latt2}) is therefore of the form
\beq
\left \{ \, - \, \bar{G} \, \Delta_s \, - \, { 1 \over \bar{G} } \;
{}^3 \! R (s) \, + \, 1 \, \right \}
\;  \Psi [ \, s \, ] \; = \; 0 \;\; ,
\label{eq:wd-latt3}
\eeq
with $\Delta_s$ a discretized form of the covariant
super-Laplacian, acting locally 
on the function space of the $s=l^2$ variables.

We shall not discuss the lattice implementation of the diffeomorphism (or momentum)
constraint in Eq.~(\ref{eq:wd-2b}) .
It can be argued that this will be satisfied automatically for a regular or random homogeneous
lattice.
This will indeed be the case for the examples we will be discussing below.


\vskip 40pt
\section{Explicit Setup for the Lattice Wheeler-DeWitt Equation}
\label{sec:calc}

In this section, we shall establish our notation and derive the
relevant terms in the discrete Wheeler-DeWitt equation for a simplex.

\subsection{2+1 dimensions}

The basic simplex in this case is of course a triangle, with vertices
and squared edge lengths labelled as in Figure 1. We set $l_{01}^2=a, \,
l_{12}^2=b, \, l_{02}^2=c$.

\vskip 10pt

\begin{center}
\epsfxsize=7cm
\epsfbox{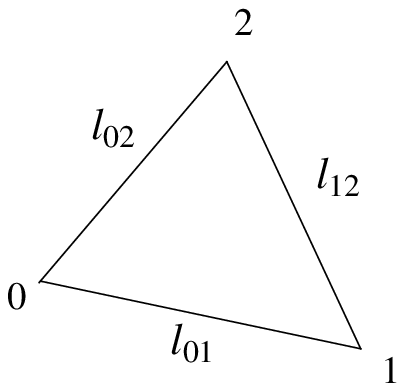}
\end{center}

\noindent{\small Figure 1. A triangle with labels.}

\vskip 10pt

The components of
the metric for coordinates based at vertex $0$, with axes along the
$01$ and $02$ edges, are
\beq
g_{11} \; = \; a , \;\;\;\;\; g_{12} \; = \; \frac{1} {2} (a + c - b),
\;\;\;\; g_{22} \; = \; c .
\eeq
The area $A$ of the triangle is given by
\beq
A^2 \; = \; \frac {1} {16} [ 2 ( ab + bc + ca ) - a^2 - b^2 - c^2 ] \; ,
\eeq
so the supermetric $G^{ij}$, according to Eq.~(\ref{eq:inverselasup}),
is
\beq
G^{ij} \; = \; \frac {1} {4A} \;\;
\left (  \matrix{
1 & -1 & -1 \cr
-1 & 1 & -1 \cr
-1 & -1 & 1
}  \right  ) ,
\eeq
with inverse
\beq
G_{ij} \; = \; - 2A \;\;
\left (  \matrix{
0 & 1 & 1 \cr
1 & 0 & 1 \cr
1 & 1 & 0
}  \right  ) .
\eeq
Thus for the triangle we have
\beq
G_{ij} \; {\partial^2 \over \partial s_i \; \partial s_j} \; = \; - 4 A
\left ( {\partial^2 \over \partial a \; \partial b} \; + \; {\partial^2
\over \partial b \; \partial c} \; + \; {\partial^2 \over \partial c \;
\partial a} \right ),
\eeq
and the Wheeler-DeWitt equation is  
\beq
\left  \{  (16 \pi G)^2 \; 4 A \; \left ( {\partial^2 \over \partial a
\; \partial b} \; + \; {\partial^2 \over \partial b \; \partial c} \;
+ \; {\partial^2 \over \partial c \; \partial a} \right ) \; - \;
2 \; n_{\sigma h} \, \sum_h \delta_h \; + \; 2 \lambda A \right \} \Psi
[\, s \,] \; = \; 0,
\label{eq:wd-2d}
\eeq
where the sum is over the three vertices $h$ of the triangle. 
The combinatorial factor $ n_{\sigma h} $ ensures the correct normalization
for the curvature term, since the latter has to give the lattice version of
$ \int \sqrt{g} \; {}^2 R = 2 \sum_h \delta_h $  when summed over all simplices
(triangles in this case) $\sigma$.
The inverse of $ n_{\sigma h} $ counts therefore the number of times
the same vertex appears in various neighboring triangles, and consequently depends 
on the specific choice of underlying lattice structure.

Alternatively, we can evaluate $G_{ij,kl} \; {\partial^2 \over
\partial g_{ij} \; \partial g_{kl}}$ directly, using
\beq
G_{ij,kl} \; = \; \frac {1} {2 \, \sqrt{g}} ( g_{ik} \, g_{jl} \; + \;
g_{il} \, g_{jk} \; - \; 2 \, g_{ij} \, g_{kl})
\eeq
(note the different coefficient of the last term in two dimensions),
with the metric $g_{ij}$ as found above. The derivatives with respect
to the metric are expressed in terms of derivatives with respect to
squared edge lengths by 
\beq
{ \partial \over \partial \, g_{ij} (s) } \; = \; \sum_m  
{ \partial \, s_m  \over \partial \, g_{ij} }
{ \partial \over \partial \, s_m  } \; .
\eeq
This leads to 
\beq
{\partial \over \partial g_{11}} \; = \; {\partial \over \partial a}
\; + \; {\partial \over \partial b} ,
\eeq
\beq
{\partial \over \partial g_{12}} \; = \; {\partial \over \partial
g_{21}} \; = \; - \; {\partial \over \partial b}
\eeq
and
\beq
{\partial \over \partial g_{22}} \; = \; {\partial \over \partial b} \;
+ \; {\partial \over \partial c}.
\eeq
This procedure gives exactly the same expression for the kinetic term.

\vskip 10pt

\begin{center}
\epsfxsize=7cm
\epsfbox{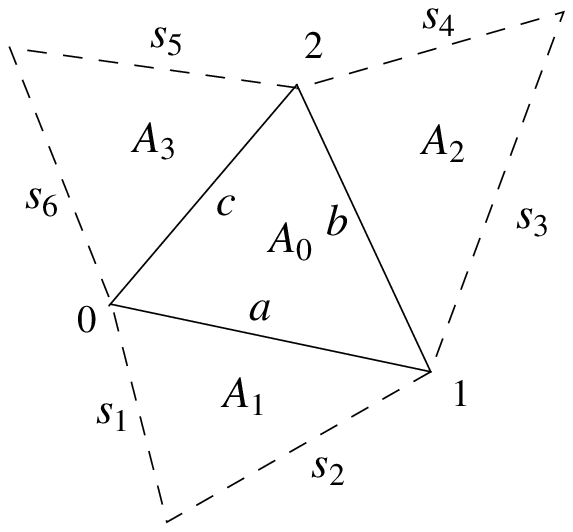}
\end{center}

\noindent{\small Figure 2. Neighbors of a given triangle.The above picture 
is supposed to illustrate the fact that the Laplacian $\Delta_{l^2}$ appearing 
in the kinetic term of the lattice Wheeler-DeWitt equation 
(here in 2+1 dimensions) contains edges $a,b,c$ that belong both to the 
triangle in question, as well as to several neighboring triangles (here three of them) 
with squared edges denoted sequentially by $s_1 =l_1^2 \dots s_6=l_6^2 $.}

\vskip 20pt

\subsection{3+1 dimensions}

In this case, both methods described  for 2+1 dimensions can be
followed, but one is much easier than the other.

\vskip 10pt

\begin{center}
\epsfxsize=7cm
\epsfbox{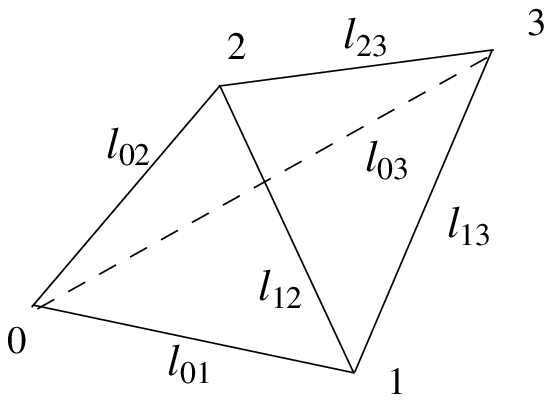}
\end{center}

\noindent{\small Figure 3. A tetrahedron with labels.}

\vskip 10pt

For ease of notation, we define $l_{01}^2=a, \, l_{12}^2=b, \, l_{02}^2=c, \,
l_{03}^2=d, \, l_{13}^2=e, \, l_{23}^2=f$.
For the tetrahedron labelled as in Figure 3, we have
\beq 
g_{11} \; = \; a \, , \;\;\;\; g_{22} \; = \; c \, , \;\;\; g_{33} \;
= \; d \, ,
\eeq
\beq
g_{12} \; = \; \frac {1} {2} (a \, + \, c \, - \, b) \, , \;\;\;
g_{13} \; = \; \frac {1} {2} (a \, + \, d \, - \, e) \, , \;\;\;
g_{23} \; = \; \frac {1} {2} (c \, + \, d \, - \, f) \; ,
\eeq
and its volume $V$ is given by 
\bea
V^2 & = & \; \frac {1} {144} [ \, af (-a-f+b+c+d+e) 
\, + \, bd (-b-d+a+c+e+f) \, + \, 
\nonumber \\
&\;\;\;\; + & ce (-c-e+a+b+d+f) \, - \, abc \, - \, ade \, - \, bef \, - \, cdf ] \; .
\eea
The matrix $G^{ij}$ is then given by
\beq
G^{ij} \; = \; - \; \frac {1} {24 V}
\left (  \matrix{
-2f & e+f-b & b+f-e & d+f-c & c+f-d & p \cr
e+f-b & -2e & b+e-f & d+e-a & q & a+e-d \cr
b+f-e & b+e-f & -2b & r & b+c-a & a+b-c \cr
d+f-c & d+e-a & r & -2d & c+d-f & a+d-e \cr
c+f-d & q & b+c-a & c+d-f & -2c & a+c-b \cr
p & a+e-d & a+b-c & a+d-e & a+c-b & -2a 
}  \right ),
\eeq
where 
\beq
p = -2a-2f+b+c+d+e, \;\;\; q = -2c-2e+a+b+d+f, \;\;\; r = -2b-2d+a+c+e+f.
\eeq
It is nontrivial to invert this (although it can be done), so instead of using $G_{ij} \;
{\partial^2 \over \partial s_i \partial s_j}$, we evaluate
\beq
G_{ij,kl} \; = \; \frac {1} {2 \sqrt{g}} (g_{ik} \, g_{jl} \; + \;
g_{il} \, g_{jk} \; - \; g_{ij} \, g_{kl}),
\eeq
with
\bea
{ \partial \over \partial \, g_{11} } & = & 
{ \partial \over \partial \, a  }
\, + \, 
{ \partial \over \partial \, b  }
\, + \, 
{ \partial \over \partial \, e  }
\nonumber \\
{ \partial \over \partial \, g_{22} } & = & 
{ \partial \over \partial \, b  }
\, + \, 
{ \partial \over \partial \, c  }
\, + \, 
{ \partial \over \partial \, f  }
\nonumber \\
{ \partial \over \partial \, g_{33} } & = & 
{ \partial \over \partial \, d  }
\, + \, 
{ \partial \over \partial \, e  }
\, + \, 
{ \partial \over \partial \, f  }
\nonumber \\
{ \partial \over \partial \, g_{12} } & = & 
- \, { \partial \over \partial \, b  }
\nonumber \\
{ \partial \over \partial \, g_{13} } & = & 
- \, { \partial \over \partial \, e  }
\nonumber \\
{ \partial \over \partial \, g_{23} } & = & 
- \, { \partial \over \partial \, f  }
\eea
The matrix representing the coefficients of the derivatives with
respect to the squared edge lengths is given in the Appendix, and is
the inverse of $G^{ij}$ found earlier. This is a nontrivial result as
it acts as confirmation of the Lund-Regge expression which was derived
in a completely different way.

Then in 3+1 dimensions, the discrete Wheeler-DeWitt equation is 
\beq
\left \{ \, - \, (16 \pi G)^2 \, G_{ij} {\partial^2 \over
\partial s_i \partial s_j} \; - \; 2 \, n_{\sigma h} \, \sum_{h}
\sqrt{s_h} \, \delta_h \; + \; 2 \lambda V \right \} \Psi [ \, s \, ] \;
= \; 0,
\label{eq:wd-3d}
\eeq
where the sum is over hinges (edges) $h$ in the tetrahedron.
Note the mild nonlocality of the equation in that the curvature term,
through the deficit angles, involves edge lengths from neighboring tetrahedra.
In the continuum, the derivatives also give some mild nonlocality.

The discrete Wheeler-DeWitt equation is hard to solve analytically,
even in 2+1 dimensions, because of the complicated dependence on
edge lengths in the curvature term, which involves arcsin or arccos of
convoluted expressions. When the curvature term is negligible, the
differential operators may be transformed into derivatives with
respect to the area (in 2+1 dimensions)  or the volume (in
3+1 dimensions) and solutions found for the wave function,
$\Psi$. 
Figures 4 and 5 give a pictorial representation of lattices that can be used for 
numerical studies of quantum gravity in 2+1 and 3+1 dimensions, respectively.

\vskip 10pt

\begin{center}
\epsfxsize=7cm
\epsfbox{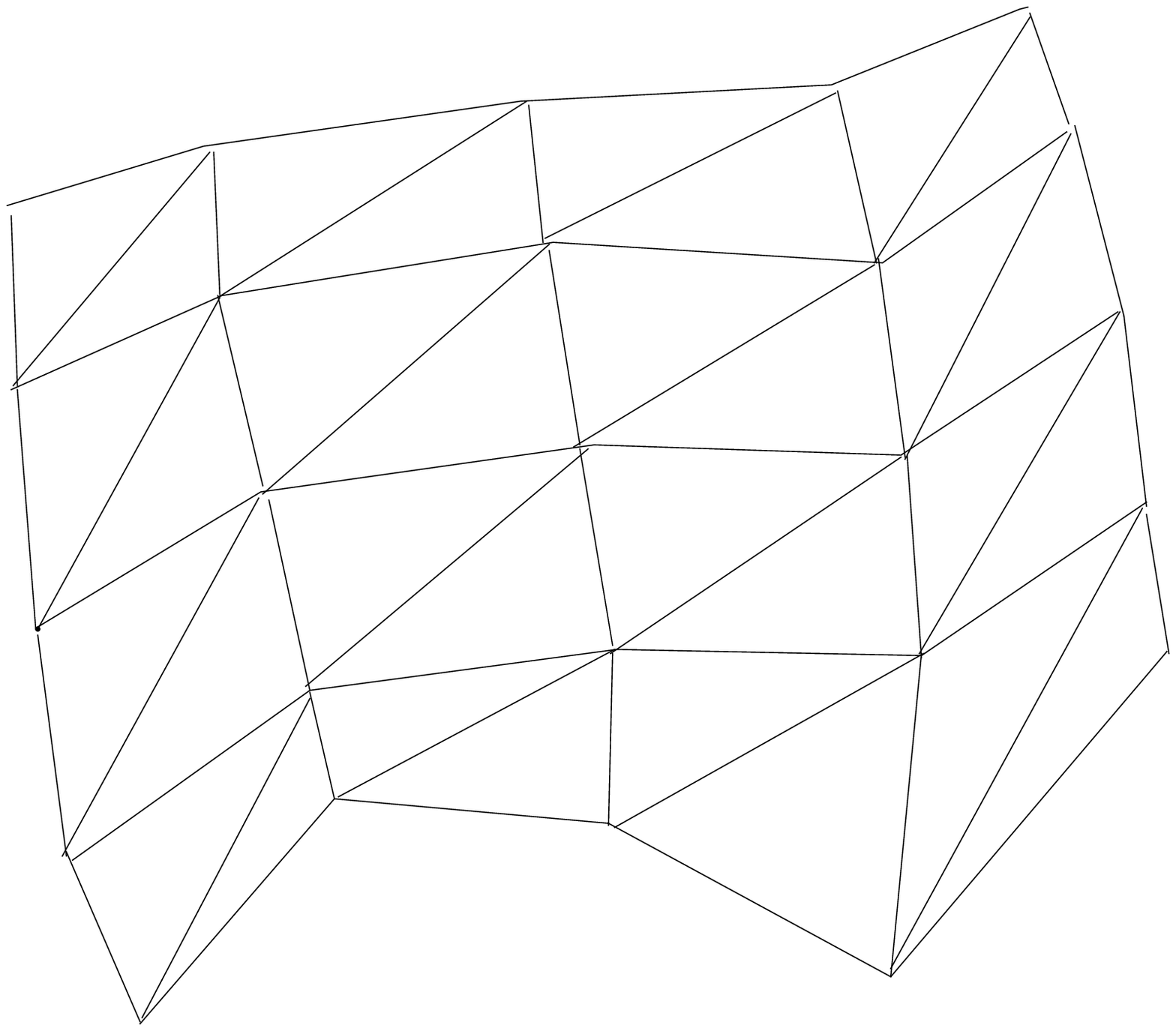}
\end{center}

\noindent{\small Figure 4. A small section of a suitable spatial lattice for quantum gravity in 2+1 dimensions.}

\vskip 10pt

\vskip 10pt

\begin{center}
\epsfxsize=7cm
\epsfbox{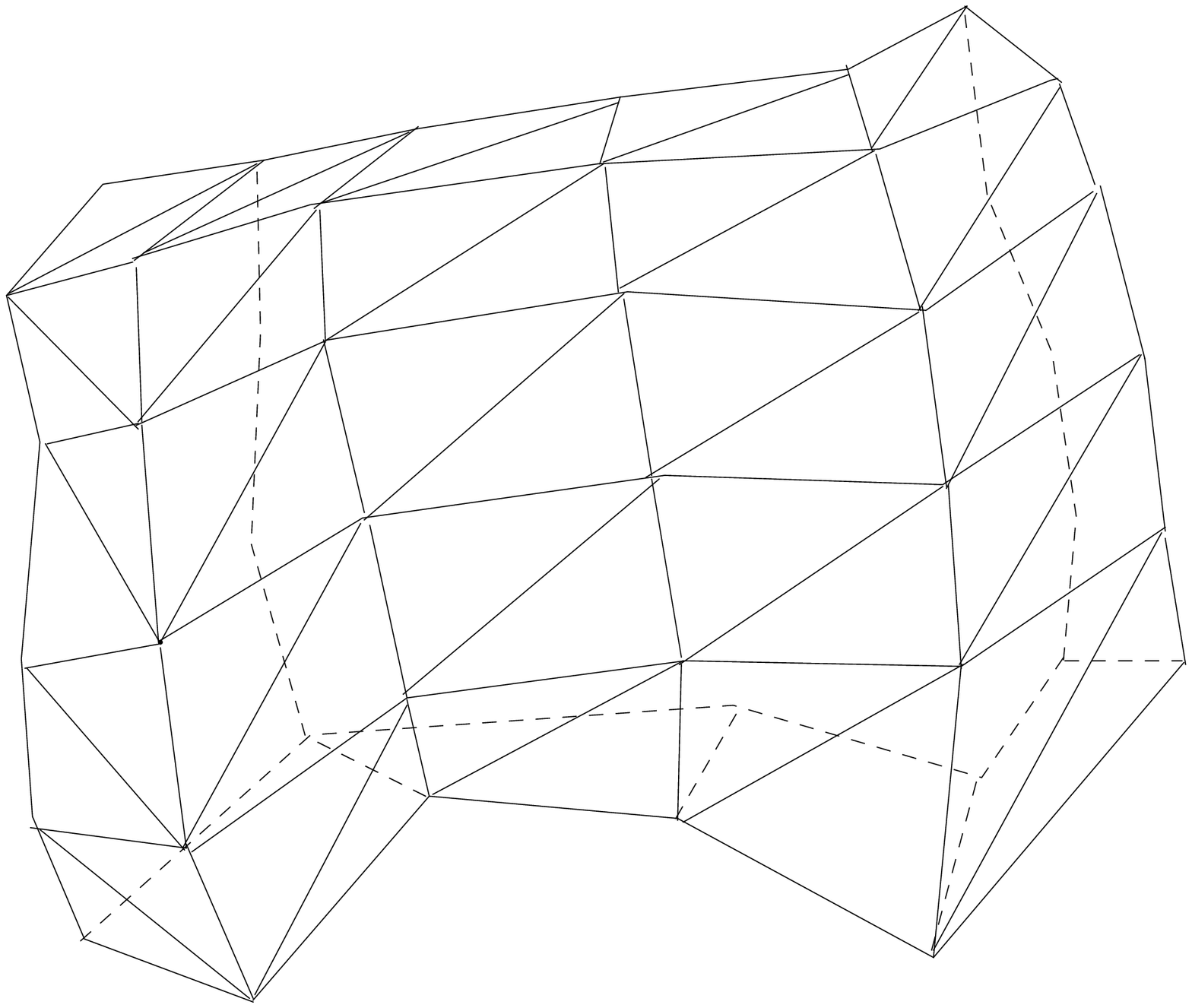}
\end{center}

\noindent{\small Figure 5. A small section of a suitable spatial lattice for quantum gravity in 3+1 dimensions.}

\vskip 10pt

\subsection{Solution of the triangle problem in 2+1 dimensions}
\label{sec:triangle}

In this section we will consider the solution of the Wheeler-DeWitt equation
for a single triangle.
The present calculation is a necessary starting point and should provide a basic stepping
stone for the strong coupling expansion in $1/G$.
In addition it will show the physical nature of the wavefunction solution deep in
the strong coupling regime.
Note that for $1/G \rightarrow 0 $ the coupling term between different simplices,
which is due to the curvature term,
disappears and one ends up with a completely decoupled problem, where
the edge lengths in each simplex fluctuate independently.
This is of course quite analogous to what happens in gauge theories on the lattice
at strong coupling, the chromo-electric field fluctuates independently on each link,
giving rise to short range correlations, a mass gap and confinement.
Here it is this single-simplex probability amplitude that we will set out to compute.

As in the Euclidean lattice theory of gravity, it will be convenient to factor out an
overall scale from the problem, and set the (un-scaled) cosmological constant
equal to one \cite{hw84}  (see Table II).
Recall that the Euclidean path integral weight contains a factor $ P(V) \propto \exp (- \lambda_0 V) $ 
where $V = \int \sqrt{g} $ is the total volume on the lattice.
The choice $\lambda_0=1$ then fixes this overall scale once and for all.
Since $\lambda_0 = 2 \lambda / 16 \pi G $ one then has $\lambda = 8 \pi G  $
in this system of units.
In the following we will also find it rather convenient to introduce the scaled coupling $\tilde \lambda$
\beq
\tilde \lambda \; \equiv \; { \lambda \over 2 } \left ( { 1 \over 16 \pi G } \right )^2
\label{eq:tilde}
\eeq
so that for $\lambda_0=1 $ (in units of the UV cutoff, or of the fundamental lattice spacing)
one has $\tilde \lambda = 1/ 64 \pi G $.

Moreover, it will often turn out to be desirable to avoid large numbers of factors of $16 \pi$'s
by the replacement, which we will follow from now on in this section, of $ 16 \pi G \rightarrow G $.
Then $\tilde \lambda = 1 / 4 G $ is the natural expansion parameter.
Note that the coupling $ \tilde \lambda$  has dimensions of length to the minus four,
or inverse area squared, in 2+1 dimension, and length to the minus six, or inverse
volume squared, in 3+1 dimensions.

Now, from Eq.~(\ref{eq:wd-2d}), the Wheeler-DeWitt equation for a single 
triangle and constant curvature density $R$ reads
\beq
\left  \{  \, (16 \pi G)^2 \; 4 A_\Delta \; \left ( {\partial^2 \over \partial a
\; \partial b} \; + \; {\partial^2 \over \partial b \; \partial c} \;
+ \; {\partial^2 \over \partial c \; \partial a} \right )
\; + \; ( 2 \lambda - R )\, A_\Delta \right \} \Psi
[\, s \,] \; = \; 0,
\eeq
where $a,b,c$ are the three squared edge lengths for the given triangle, and 
$A_\Delta$ is the area of the same triangle.
In the following we will take for simplicity $R=0$.
Equivalently one needs to solve
\beq
\left  \{  
{\partial^2 \over \partial a
\; \partial b} \; + \; {\partial^2 \over \partial b \; \partial c} \;
+ \; {\partial^2 \over \partial c \; \partial a}
\; + \; \tilde \lambda 
\right \} \Psi [\, a,b,c \,] \; = \; 0 \; .
\eeq
If one sets 
\beq
\Psi[ \, s \, ] \; = \; \Phi[ \, A_\Delta \, ],
\eeq
then one can show that
\beq
{\partial^2 \over \partial a \; \partial b} \; \Psi \; = \; { 1 \over (16 A_\Delta)^2 } 
\; (b \, + \, c \, - \, a) \; (a \, + \, c \, - \, b) \;
\left ( { d^2 \Phi \over dA_\Delta ^2 } \; - \; 
\frac {1} {A_\Delta } { d \Phi \over dA_\Delta } \right ) \; + \; \frac {1}
{16 A_\Delta } \; { d \Phi \over dA_\Delta } \; .
\eeq
Summing the partial derivatives leads to the equation
\beq
A_\Delta \; { d^2 \Phi \over dA_\Delta ^2 } \; + \; 2 \; { d \Phi \over dA_\Delta }  
\; + \; 16 \; \tilde \lambda \; A_\Delta \; \Phi \; = \; 0 \; .
\eeq
Solutions to the above equation are given by
\beq
\Psi [\, a,b,c \,] \; = \; {\rm const.}  \; {1 \over A_\Delta } 
\exp \left [ \, \pm \, i \cdot 4 A_\Delta \sqrt{ \tilde \lambda} \; \right ] \; ,
\eeq
or alternatively by
\beq
\Psi [\, a,b,c \,] \; = \; { 1 \over A_\Delta } 
\left [ \, c_1 \cos \left ( 4 A_\Delta \sqrt{ \tilde \lambda}  \, \right ) \, + \, 
c_2  \sin \left  ( 4 A_\Delta \sqrt{ \tilde \lambda}  \, \right )  \right ] \; .
\eeq
Note the remarkable, but not entirely unexpected, result that the wavefunction 
only depends on the area of the triangle $A_\Delta  (a,b,c) $.
In other words, it depends on the geometry only.
Regularity of the wavefunction as the area of the triangle approaches
zero, $A_\Delta \rightarrow 0 $, requires for the constant $c_1=0$.
Therefore the correct quantum-mechanical solution is unambiguously determined,
\beq
\Psi [\, a,b,c \,] \; = \; { 1 \over \sqrt{ 2 \pi \sqrt{\tilde \lambda} } } \;
 { 1 \over A_\Delta } \; 
\sin \left ( 4 A_\Delta \sqrt{ \tilde \lambda}  \, \right ) \; .
\eeq
The overall normalization constant has been fixed by
the standard rule of quantum mechanics,
\beq
\int_0^\infty  d A_\Delta \; | \, \Psi (A_\Delta ) \, |^2 \; = \; 1 \; .
\eeq
Moreover we note that a bare $\lambda <0 $ is not possible, and that 
the oscillatory nature of the wavefunction is seen here to give rise to well-defined
peaks in the probability distribution for the triangle area, located at
\beq
\left ( A_\Delta \right )_n \; = \; { n \, \pi \over 4 \, \sqrt{ \tilde \lambda} }
\eeq
with $n$ integer.

\subsection{Solution of the tetrahedron problem in 3+1 dimensions}
\label{sec:tetrahedron}

In this section we will consider the nature of quantum-mechanical
solutions for a single tetrahedron.
Now, from Eq.~(\ref{eq:wd-3d}), the Wheeler-DeWitt equation for a single tetrahedron
with a constant curvature density term $R$ reads
\beq
\left \{ \, - \, (16 \pi G)^2 \, G_{ij} {\partial^2 \over
\partial s_i \partial s_j} 
\;  + \; (2 \lambda - R) \, V \right \} \Psi [ \, s \, ] \;
= \; 0,
\eeq
where now the squared edge lengths $s_1 \dots s_6 $ are all part of the
same tetrahedron, and $G_{ij}$ is given by a rather complicated, but explicit,
$ 6 \times 6 $ matrix given earlier.

As in the 2+1 case discussed in the previous section, here too it is
found that, when acting on functions of the tetrahedron volume,
the Laplacian term still returns some other function of the volume only,
which makes it possible to readily obtain a full solution for the wavefunction.
In terms of the volume of the tetrahedron $V_T$ one has the
equivalent equation for $\Psi [s]=f (V_T)$ (we again replace $16 \pi G \rightarrow G$ 
from now on)
\beq
{7 \over 16 } G \, f' (V_T ) + { 1 \over 16 } G \, V_T \, f'' (V_T ) + 
{1 \over G } ( 2 \lambda - R ) \, V_T \, f (V_T )  \; = \; 0
\eeq
with primes indicating derivatives with respect to $V_T$.
From now on we will set the constant curvature density $R$=0; 
then the solutions are Bessel functions $J_m$ or $Y_m$ with  $m=3$,
\beq
\psi_R ( V_T ) \; = \; {\rm const. } \; 
J_3 \left ( 4 \sqrt{2}  \, { \sqrt{\lambda} \over G }  V_T \right ) / V_T^3
\eeq
or
\beq
\psi_S ( V_T ) \; = \; {\rm const. } \; 
Y_3 (\left ( 4 \sqrt{2}  \, { \sqrt{\lambda} \over G }  V_T \right ) / V_T^3 \; .
\eeq
Only $J_m (x)$ is regular as $x \rightarrow 0$, 
$ J_m (x) \sim \Gamma (m+1)^{-1} (x/2)^m $.
So the only physically acceptable wavefunction is
\beq
\Psi (a,b, \dots f ) \; = \; \Psi ( V_T ) \; = \; {\cal N } \; 
{ J_3 (\left ( 4 \sqrt{2}  { \sqrt{\lambda} \over G }  V_T \right ) 
\over  V_T^3 }
\eeq
with the normalization constant $\cal N$ given by
\beq
{\cal N} \; = \; 
{ 45 \sqrt{77 \pi} \over 1024 \, 2^{3/4} } \; 
\left ( { G \over \sqrt{\lambda} } \right )^{5/2} \; .
\eeq
The latter is obtained from the wavefunction normalization requirement
\beq
\int_0^\infty d V_T \, | \, \Psi ( V_T ) \, |^2 \; = \; 1 \; .
\eeq
Consequently the average volume of a tetrahedron is given by
\beq
\langle \; V_T \; \rangle \; \equiv \; \int_0^\infty d V_T \cdot 
V_T \cdot | \, \Psi ( V_T ) \, |^2
\; = \; {31185 \pi G \over 262144 \sqrt{2} \, \sqrt{\lambda}  }
\; = \; 0.2643 \; {G \over \sqrt{\lambda}  } \; .
\eeq
This last result allows us to define an average lattice spacing,
by comparing it to the value for an equilateral tetrahedron
which is $V_T = (1/ 6 \sqrt{2} ) \, l_0^3 $.
One then obtains for the average lattice spacing at strong coupling
\beq
l_0 \; = \; 1.3089 \; \left ( { G \over \sqrt{\lambda}  } \right )^{1/3} \; . 
\eeq
Note that in terms of the parameter $\tilde \lambda$ defined in Eq.~(\ref{eq:tilde})
one has in all the above expressions $\sqrt{\lambda}/G = \sqrt{ 2 \tilde \lambda}$.

The above results further show that for strong gravitational 
coupling, $1/G \rightarrow 0$, lattice quantum gravity has a finite correlation
length, of the order of one lattice spacing,
\beq
\xi \; \sim l_0 \; .
\eeq
This last result is simply a reflection of the fact that for large $G$
the edge lengths, and therefore the metric, fluctuate more or less 
independently in different spatial regions due to the absence of the curvature term.
The same is true in the Euclidean lattice theory as well, in the same limit \cite{hw84}.
It is the inclusion of the curvature term that later leads to a coupling
of fluctuations between different spatial regions.
Only at the critical point in $G$, if one can be found, is the correlation length,
measured in units of the fundamental lattice spacing, expected to diverge \cite{hbook}.
This last circumstance should then allow the construction of a proper lattice 
continuum limit, as is done in the Euclidean lattice theory of gravity \cite{ham00}
(and in many other lattice field theories as well).

\vskip 40pt
\section{Perturbation Theory in the Curvature Term}
\label{sec:perturbation}

As shown in the previous section, in a number of instances 
it is not difficult to find the solution $\Psi$
of the Wheeler-DeWitt equation in the strong coupling (large $G$) limit, where the
curvature term is neglected, and only the kinetic and $\lambda$ terms are retained.
Then the dynamics at different points decouples, and the wavefunction can be
written as a product of relatively simple wavefunctions.
It is then possible, at least in principle, to include the curvature term as a perturbation to
the zeroth order solution.
Accordingly, the unperturturbed Wheeler-DeWitt Hamiltonian is denoted by $H_0$ 
\beq
H_0 \; \equiv \; -  \, 16 \pi G \cdot G_{ij} ( l^2 ) \, 
{ \partial^2 \over \partial l^2_{i} \, \partial l^2_{j}  }
\; + \; { 1 \over 16 \pi G  } \, \sqrt{g (l^2) } \; \cdot 2 \lambda 
\label{eq:h0}
\eeq
and the perturbation by $H_1$
\beq
H_1 \; \equiv \;  - \; { 1 \over 16 \pi G  } \, \sqrt{g (l^2) } \; {}^3 \! R (l^2) \; .
\label{eq:h1}
\eeq
The corresponding  unperturbed wavefunction is denoted by $\Psi_0$, and
satisfies
\beq
H_0 \; \Psi_0 \; = \; 0 \; .
\eeq
To the next order in Raleigh- Schr\"odinger perturbation theory one needs to solve
\beq
\left ( H_0 + \epsilon \, H_1 \right ) \; \Psi \; = \; 0
\eeq
where for $\Psi$ one sets as well
\beq
\Psi \; = \; \Psi_0 \; \exp \left \{  \epsilon \, \Psi_1 \right \} \; .
\eeq
The sought-after first order correction $\Psi_1 $ is then given by the solution of
\beq
H_0 ( \Psi_0 \, \Psi_1 ) + H_1 \, \Psi_0 \; = \; 0 \; .
\eeq
Higher order corrections can then be obtained in analogous fashion.
It would seem natural to search for a solution (here specifically in 3+1 dimensions)
of the form
\beq
\Psi \; \sim \; \exp \left \{ - \, \alpha ( \lambda, G ) \, \sum_\sigma \, V_\sigma 
\, + \, \beta ( \lambda, G ) \, \sum_h \delta_h \, l_h + \dots \right \}
\label{eq:wf_exp}
\eeq
with $\alpha$ and $\beta$ given by power series 
\bea
\alpha ( \lambda, G )
 & = & { \sqrt{\lambda} \over G } \; \sum_{n=0}^\infty  \, \alpha_n \, ( G \lambda )^n
\nonumber \\
\beta ( \lambda, G )
& = & \left (  { \sqrt{\lambda} \over G } \right )^{1/3} \; 
\sum_{n=0}^\infty  \, \beta_n \, ( G \lambda )^n \; .
\eea
The dots in Eq.~(\ref{eq:wf_exp}) indicate possible higher derivative terms in the exponent of
the wavefunction.

\vskip 40pt
\section{Variational Method for the Wavefunction $\Psi$}
\label{sec:variational}

In this section we will describe some simple applications of the variational method for
quantum gravity, based on the lattice Wheeler-DeWitt equation proposed earlier.
The power of the variational method is well known and appreciated in 
nonrelativistic quantum mechanics, atomic physics, and many other physically relevant applications.
Its success generally rests on the ability of finding a suitable, often physically motivated, 
wavefunction with the lowest possible energy, thereby providing an approximation
to both the ground state energy and the ground state wavefunction.
In practice the wavefunction is often written as some sort of product of orbitals, dependent on
a number of suitable parameters, which are later determined by minimization.

Here we will write therefore an ansatz for the variational wavefunction, dependent on a number of
free variational parameters
\beq
\Psi [l^2 ] \; = \; \Psi [l^2 ; \alpha , \beta , \gamma \dots ] \; ,
\eeq
and later require that the resulting wavefunction either satisfy the Wheeler-DeWitt
equation, or that its energy functional 
\beq
E ( \alpha , \beta , \gamma \dots ) \; = \; 
{ 
\langle \Psi [ \, l^2 \, ] \, |
\left \{ \, -  \, 16 \pi G \cdot G_{ij} ( l^2 ) \, 
{ \partial^2 \over \partial l^2_{i} \, \partial l^2_{j}  }
\; - \; { 1 \over 16 \pi G } \, \sqrt{g (l^2) } \; \left [ \; 
{}^3 \! R (l^2) \, - \, 2 \lambda \; \right ] \; \right \} \; 
| \Psi [ \, l^2 \, ] \rangle 
\over 
\langle \Psi [ \, l^2 \, ] \, | \Psi [ \, l^2 \, ] \rangle 
}
\label{eq:en-variational}
\eeq
be as close to zero as possible, $ |E|^2 = $ min.
This procedure should then provide a useful algebraic relation between the variational
parameters, and thus allow their determination.
\footnote{The continuum analog of the above expression would have the following general structure:
\beq
E \; = \; { \int d^3 {\bf x} \, \int [d g] \; \Psi^* [g] \cdot 
\left [ - G \, \Delta_g  - {1 \over G } \, \sqrt{g}  \, (R - 2 \lambda ) \right ] \cdot \Psi [g]
\over 
\, \int [d g] \; \Psi^* [g] \cdot \Psi [g] } \; .
\eeq
Similar energy functionals were considered some time ago by Feynman in his variational 
study of Yang-Mills theory in 2+1 dimensions \cite{fey81}.
The main difference with gauge theories is that here the Hamiltonian contains two
terms (kinetic and curvature terms) that enter with {\it opposite} signs, whereas
in the gauge theory case both terms 
(the ${\bf E}^2$ term and the $ (\nabla \times {\bf A})^2$ term) 
just add to each other. 
Feynman then argues that in the gauge theory the state of lowest energy corresponds 
necessarily to a minimum for both contributions.
}

Here we will consider the following correlated product variational wavefunction (in general dimension)
\beq
\Psi [ l^2 ] \, = \,  
Z^{-1/2} \; e^{- \alpha \sum_\sigma V_\sigma + \beta \sum_h \delta_h V_h  + \dots  }
\nonumber \\
\, = \,  Z^{-1/2} \; \prod_\sigma  \left ( e^{- \alpha \, V_\sigma } \right ) \; 
\prod_h  \left ( e^{ \beta \, \delta_h V_h  } \right ) \, \times \dots
\label{eq:wave}
\eeq
with variational parameters $ \alpha, \beta, \dots $ real or complex.
Here the $\sum_\sigma V_\sigma$ is the usual volume term in $d$ dimensions,
and $\sum_h \delta_h V_h$ the usual Regge curvature term, in the same number
of dimensions. 
The dots indicate possible additional contributions, perhaps
in the form of invariant curvature squared terms. 
In the atomic physics literature these types of product wavefunctions 
are sometimes known as Jastrow-Slater wavefunctions \cite{car80,sur82}.
Note that the above wavefunction is very different from the ones used in
minisuperspace models, as it still depends on infinitely many lattice degrees of
freedom in the thermodynamic limit (the limit in which the number of lattice 
sites is taken to infinity).

The wavefunction normalization constant $Z ( \alpha , \beta , \gamma \dots ) $ is given by 
\beq
Z \; = \; \int [ d l^2 ] \; | \, \Psi [ l^2; \alpha, \beta, \dots\,  ] \, |^2 
\; = \;  \int [ d l^2 ] \; \exp \left \{
- 2 Re \alpha \sum_\sigma V_\sigma + 2 Re \beta \sum_h \delta_h V_h  + \dots  
\right \}
\label{eq:zdef}
\eeq
and represents the partition function for Euclidean lattice quantum gravity, but 
{\it in one dimension less}.
One would expect at least $Re \alpha >0 $ to ensure convergence of the path
integral; the trick we shall employ below is to obtain the relevant averages
by analytic continuation in $\alpha$ and $\beta$ of the corresponding
averages in the Euclidean theory (for which $\alpha $ and $\beta$ are real). 
Here the expression $ [d l^2 ] $ is the usual integration measure over the edge 
lengths \cite{hw99}, a lattice version of the DeWitt invariant functional measure 
over continuum metrics $[d g_{\mu\nu} ] $.
The definition of $Z$ requires that the functional integral in Eq.~(\ref{eq:zdef}) actually
exists, which might or might not require some suitable regularization: for example
by the addition of curvature squared terms whose amplitude is sent to zero
at the end of the calculation.

Next one needs to compute the expectation value
\beq
\langle \, \Psi [ \, l^2 \, ] \, | \, H \, | \, \Psi [ \, l^2 \, ] \, \rangle 
\eeq
with 
\beq
H  \; \equiv \; 
-  \, 16 \pi G \cdot G_{ij} ( l^2 ) \, 
{ \partial^2 \over \partial l^2_{i} \, \partial l^2_{j}  }
\; - \; { 1 \over 16 \pi G } \, \sqrt{g (l^2) } \; \left [ \; 
{}^3 \! R (l^2) \, - \, 2 \lambda \; \right ]
\label{eq:h-variational}
\eeq
which in turn is made up of three contributions, each of which
can be evaluated separately.
In terms of explicit lattice averages, one needs the three averages, or
expectation values,
\beq
\langle \, \Psi [ \, \alpha, \beta, \dots \, ] \, |
\left \{ \, - \sum_\sigma \Delta_{l^2}  ( \sigma )\right \} 
| \, \Psi [ \, \alpha, \beta, \dots \, ] \, \rangle 
\label{eq:en-ke}
\eeq

\beq
\langle \, \Psi [ \, \alpha, \beta, \dots \, ] \, |
\left \{ \, \sum_\sigma V_\sigma \, \right \} \; 
| \, \Psi [ \, \alpha, \beta , \dots \, ] \, \rangle 
\label{eq:en-vol}
\eeq

\beq
\langle \, \Psi [ \, \alpha, \beta, \dots \, ] \, |
\left \{ \, 2 \sum_h \delta_h l_h  \, \right \} \; 
| \, \Psi [ \, \alpha, \beta , \dots \, ] \, \rangle  \;
\label{eq:en-r}
\eeq
with
\beq
\Delta_{l^2}  ( \sigma ) \; \equiv \; 
G_{ij} ( l^2 ) \, { \partial^2 \over \partial l^2_{i} \, \partial l^2_{j}  } \; .
\label{eq:laplacian_l}
\eeq
Note that we have summed over all lattice points by virtue of the assumed
homogeneity of the lattice: the local {\it average} is expected to be
the same as the average of the corresponding sum, divided by
the overall number of simplices.
Thus, for example, 
$ \langle \, \Psi \, | \, \sum_\sigma V_\sigma \, | \, \Psi \, \rangle  =
N_\sigma \, \langle \, \Psi \, | \, V_\sigma \, | \, \Psi \, \rangle $ etc.
At the same time one has, by virtue of our choice of wavefunction,
\beq
\sum_\sigma V_\sigma \, | \, \Psi [ \, \alpha, \beta, \gamma , \dots \, ] \, \rangle
\; = \; - { \partial \over \partial \alpha } \;  | \, \Psi [ \, \alpha, \beta, \dots \, ]  \, \rangle
\label{eq:der1l}
\eeq
\beq
\sum_h \delta_h l_h \, | \, \Psi [ \, \alpha, \beta, \gamma , \dots \, ] \, \rangle
\; = \; { \partial \over \partial \beta } \;  | \, \Psi [ \, \alpha, \beta, \dots \, ] \, \rangle
\eeq
and also for a given simplex labeled by $\sigma$
\beq
- \; \Delta_{l^2}  ( \sigma ) \; e^{- \alpha \, V_\sigma }  \; =  \; f ( V_\sigma ) 
\eeq
where $f$ is some known function.
More specifically in 2+1 dimensions one finds (here $A_\Delta$ is the area of the relevant triangle)
\beq
- \; \Delta_{l^2}  ( \sigma ) \; A_\Delta ^n  \; = \; { 1 \over 4 } \, n \, (n+1) \, A_\Delta ^{n-1} 
\eeq
\beq
- \; \Delta_{l^2}  ( \sigma ) \; F (A_\Delta) \; = \; { 1 \over 2 } {d F \over d A_\Delta } 
\, + \, {A_\Delta \over 4 }  { d^2 F \over d A_\Delta ^2 }
\eeq
and therefore 
\beq
- \; \Delta_{l^2}  ( \sigma ) \; e^{- \alpha A_\Delta  }  \; =  \;  
{ 1 \over 4 } \, \alpha \, ( \alpha A_\Delta - 2 ) \; e^{- \alpha A_\Delta }  \; ,
\eeq
whereas in 3+1 dimensions one has (here $V_T$ is the volume of the relevant tetrahedron)
\beq
- \; \Delta_{l^2}  ( \sigma ) \; V_T^n \; = \; { 1 \over 16 } \, n \, (n+6) \, V_T^{n-1} 
\eeq
\beq
- \; \Delta_{l^2}  ( \sigma ) \; F (V_T) \; = \; { 7 \over 16 } {d F \over d V_T } 
\, + \, {V_T \over 16 }  { d^2 F \over d V_T^2 }
\eeq
and therefore 
\beq
- \; \Delta_{l^2}  ( \sigma ) \; e^{- \alpha V_T  }  \; =  \;  
{ 1 \over 16 } \, \alpha \, ( \alpha \, V_T - 7 ) \; e^{- \alpha V_T }  \; .
\eeq
In addition, in 3+1 dimensions one needs to evaluate
\beq
- \; \Delta_{l^2}  ( \sigma ) \; e^{ \beta \sum_h l_h \delta_h  }  
\eeq
which is considerably more complicated.
Nevertheless in 2+1 dimensions the corresponding result is zero, by
the Gauss-Bonnet theorem.
One identity can be put to use to relate one set of averages to another;
it follows from the scaling properties of the lattice measure $[dl^2]$
in $d$ dimensions with curvature coupling $k=1/8 \pi G$ and unscaled
cosmological constant $\lambda_0 \equiv \lambda / 8 \pi G $ \cite{ham00}.
In three dimensions it reads
\beq
2 \, \lambda_0 \, \langle \sum_T V_T \rangle \, - \, k \, \langle \sum_h \delta_h l_h \rangle - 7 N_0 =0 
\eeq
where in the first term the sum is over all tetrahedra, and in the second term
the sum is over all hinges (edges). The quantity $N_0$ is the number of sites
on the lattice, the coefficient in front of it in general depends on the lattice coordination
number, but for a cubic lattice subdivided into simplices it is equal to 7, since there
are seven edges within each cube (three body principals, three face diagonals
and one body diagonal).
The above sum rule can then be used by making the substitution
$\lambda_0 \rightarrow 2 Re \alpha $ and
$k \rightarrow 2 Re \beta $.
In two dimensions the analogous result reads
\beq
2 \, \lambda_0 \, \langle \sum_\Delta A_\Delta \rangle 
\, - \, k \, \langle \sum_h \delta_h \rangle - 3 N_0 =0 
\eeq
with $ 2 \sum_h \delta_h = \int \sqrt{g} R = 4 \pi \chi $=const. by the Gauss-Bonnet theorem.

From now on we will focus on the 2+1 case exclusively.
In this case the curvature average of Eq.~(\ref{eq:en-r}) is very simple
\beq
\langle \int \sqrt{g} R \, \rangle \rightarrow  4 \pi \chi
\eeq
where $\chi$ is the Euler characteristic for the two-dimensional manifold.
It will also be convenient to avoid a large number of factors of
$16 \pi $'s  and make the replacement $16 \pi G \rightarrow G$ for
the rest of this section.
Putting everything together one then finds
\beq
{ E ( \alpha ) \over G  N_T } \; = \;
{1\over 4 } \, \alpha \, ( \alpha \, \bar A_\Delta -2 ) \, + \, 
{ 2 \lambda \over G^2 } \; \bar A_\Delta \, - \, {1 \over G^2 } { 4 \pi \, \chi \over N_T } \; .
\eeq
One is not done yet, since what is needed next is an estimate for the average
area of a triangle, $\bar A_\Delta $.
This quantity is given, for a general measure over edges in two dimensions of
the form $ \prod d l^2 \cdot \prod_T (A_\Delta)^\sigma $, by 
\beq
\langle A_\Delta \rangle \; = \; { 1 + {2 \over 3 } \, \sigma \over 2 \, \alpha } \; ,
\eeq
again with the requirement $Re \alpha >0 $ for the average to exist.
It will be convenient to just set in the following
$\bar A_\Delta = \langle A_\Delta \rangle \; = \; \sigma_0 / \alpha $ with 
$\sigma_0 \equiv (1 + {2 \over 3 } \, \sigma)/ 2$.
One then obtains, finally, the relatively simple result
\beq
{ E ( \alpha ) \over G N_T } \; = \;
{\sigma_0 - 2  \over 4 }  \cdot \alpha \, + \, 
{ 2 \, \lambda \, \sigma_0  \over G^2 } \cdot {1 \over \alpha } \, - \, { 4 \pi \, \chi \over G^2 N_T } \;\; .
\eeq
It would seem that, in order to avoid a potential instability, it might be safer to choose $\sigma_0 > 2$.
The roots of this equation (corresponding to the requirement 
$ \langle \, \Psi \, | \, H \, | \, \Psi \, \rangle  = 0 $ ) are given by
\beq
\alpha_0 = { 1 \over G^2 N_T ( \sigma_0 -2 ) } \left \{ 8 \pi \chi \pm \sqrt{\Delta} \right \}
\eeq
with
\beq
\Delta \equiv 64 \pi^2 \chi^2 - 8 G^2 N_T^2 \lambda \sigma_0 (\sigma_0-2) \; ,
\eeq
so that $\Delta$ is zero for
\beq
G \; = \; G_c = \pm { 2 \sqrt 2 \, \pi \, \chi \over N_T \sqrt{\lambda \, \sigma_0 ( \sigma_0 -2 ) } } \; .
\eeq
Here we select, on physical grounds, the positive root.
When $\Delta =0 $ (or $G=G_c$) the two complex roots become real, or vice-versa, with
\beq
\alpha_0 (G_c) \; = \; { N_T \lambda \, \sigma_0 \over \pi \chi } > 0 \;\;\;\;\; {\rm if} \;\; \chi  > 0 \; .
\eeq
Thus for strong coupling (large $G > G_c$) $\alpha$ is almost purely imaginary
\beq
 \alpha_0  \; = \; \pm { i \, 2 \sqrt{2 \, \lambda } \over G \, \sqrt{1-2/\sigma_0} }  \, + \,
{ 8 \pi \, \chi \over G^2 N_T \, ( \sigma_0 -2 ) } + O( 1/G^3 ) \; ,
\eeq
whereas for weak coupling (small $G < G_c$) the two roots become
\bea
\alpha_0 & \rightarrow & { N_T \lambda \, \sigma_0  \over 2 \pi \chi } + O( G^2 )
\nonumber \\
\alpha_0 & \rightarrow & { 16 \pi \chi  \over G^2 N_T (\sigma_0 -2 ) } 
- { N_T \lambda \, \sigma_0 \over 2 \pi \chi } + O( G^2 ) \; .
\eea
Note that an identical set of results would have been obtained if one
had computed $ |E (\alpha ) |^2 $ for complex alpha, and looked for minima.
This is the quantity displayed in Figures 6 and 7.

Next we come to a brief discussion of the results.
One interpretation is that the variational method, using the
proposed correlated product wavefunction in 2+1 dimensions, suggests
the presence of a phase transition for pure gravity in $G$, located at 
the critical point $G=G_c$.
This picture found here would then be in accordance with the result found in the 
{\it Euclidean}  lattice theory in Ref. \cite{hw93}, which also gave a phase transition
in three-dimensional gravity between a smooth phase (for $G>G_c$) and a branched
polymer phase (for $G<G_c$).
A similar transition was found on the lattice in four dimensions as well \cite{hw84}.
Finally, the presence of a phase transition is also inferred from continuum calculations 
for pure gravity in $\epsilon \equiv d-2 > 2 $, although the latter does not give a clear
indication on which phase is physical; nevertheless simple renormalization group
arguments suggest that the weak coupling phase describes gravitational screening,
while the strong coupling phase implies gravitational anti-screening.
This last expansion  then gives a critical point for pure gravity in 2+1 dimensions at 
$G_c = {3 \over 25} (d-2) + {45 \over 1250 } (d-2)^2 + \dots $, or $G_c \approx 0.024 $ 
in units of the cutoff \cite{aid97,eps,reu98}.
The Euclidean lattice calculation quoted earlier gives, in the same dimensions,
$ G_c \approx 0.355 $.
Note that the numerical magnitude of the critical point $G$ in lattice units, contrary
to the critical exponents, is not 
expected to be universal, and thus cannot be compared directly between
formulations utilizing different ultraviolet regulators. 
We shall not enter here into some of the known peculiarities
of three-dimensional gravity, including the absence of perturbative 
transverse-traceless radiation modes, 
and the absence of a sensible Newtonian limit; a recent discussion of these and related
issues can be found for example in \cite{hbook}, and further references cited therein.  

In 3+1 dimensions the variational calculation is quite a bit more complex, since
the integrated curvature term is no longer a constant.
In the small curvature limit and for small variational parameter $\beta$ we have obtained 
the following expansion for the variational parameter $\alpha$
\beq
\alpha_0 = \pm\,  i \, 4 \sqrt{2} \, { \sqrt{\lambda} \over G } \sqrt{ \sigma_0 \over \sigma_0 -7 } 
\, - \, { 8 \, c_0 \, \beta \over \sigma_0 -7 } \, + \, O(\beta^2 ) \; .
\eeq
Here $c_0$ is a real constant whose value we have not been able to determine yet.
The two roots are found to become degenerate and real for
\beq
G \; = \; G_c \equiv { \sqrt{ \lambda \, \sigma_0 ( \sigma_0 -7 ) } \over \sqrt{2} \, c_0 \, \beta }
\eeq
which is again suggestive of a phase transition at $G_c$ in 3+1 dimensions, as found
previously in the Euclidean theory in four dimensions \cite{hw84,ham00}.
More detailed calculations in the 3+1 case are in progress, and 
will be presented elsewhere \cite{htw11}.

We conclude this section by observing that our results suggest a 
rather intriguing relationship between the ground state wave functional of
quantum gravity in $n+1$ dimensions, and 
averages computed within the Euclidean Feynman path integral formulation in 
$n$ dimensions, i.e. in {\it one dimension less}.
Moreover, since the variational calculations presented here rely on what could be 
regarded as an improved mean field calculation, they are expected to become 
more accurate in higher dimensions, where the number of neighbors to each 
lattice point (or simplex) increases rapidly.

\vskip 20pt

\begin{center}
\epsfxsize=10cm
\epsfbox{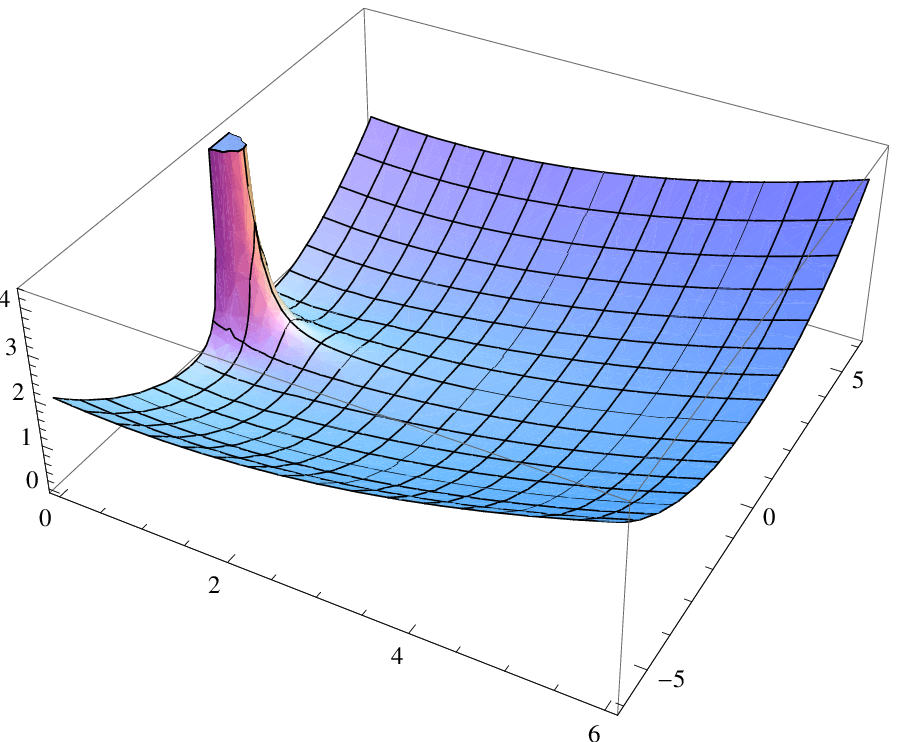}
\end{center}

\noindent{\small Figure 6. Energy surface $|E(\alpha)|^2 $ in 2+1 dimensions 
at strong coupling, $G \gg G_c$
in the ($Re \alpha$, $Im \alpha$) plane.
Note the presence of two almost purely imaginary, complex conjugate roots.
The specific values used here are $\chi=2, \, N_T=10,\, \sigma_0=3$ and $\lambda=1$.}


\vskip 20pt

\begin{center}
\epsfxsize=10cm
\epsfbox{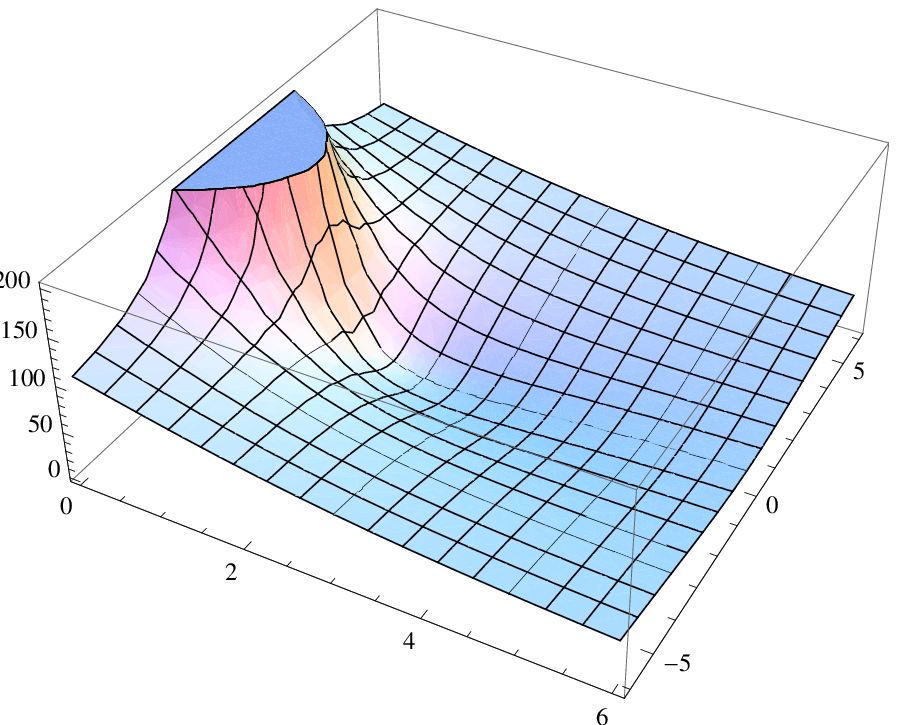}
\end{center}

\noindent{\small Figure 7. Energy surface $|E(\alpha)|^2 $  in 2+1 dimensions 
for weak coupling, $G \ll G_c$.
In this case both roots are along the real $\alpha$ axis.}

\vskip 20pt





\vskip 40pt
\section{Weak field expansion}
\label{sec:wfe}

In this section we will discuss briefly the weak field expansion for the 
proposed lattice Wheeler-DeWitt equation.
The purpose here is to show how the weak field expansion is performed,
and how results analogous to the continuum ones are obtain for
sufficiently smooth manifolds.
Such results would be of relevance to the weak coupling (small $G$) expansion,
and to an application of the $WKB$ method on the lattice, for example.
More generally a clear connection to the continuum theory, and thus between lattice
and continuum operators, is desirable, if not essential, in order to understand
the meaning of physical gravitational averages, such as average curvature etc.
First we note here that the lattice kinetic term (the one involving $G_{ij}$)
has the correct continuum limit, essentially by construction.
On the other hand the curvature term appearing in the discrete Wheeler-DeWitt equation 
in 3+1 dimension is nothing but the integrand in the Regge expression for the 
Einstein-Hilbert action in three dimensions,
\beq
I_E  \; =  \; - \, k \, \sum_ {\rm edges \; h }  l_h \delta _ h \; .
\eeq
The expansion of this action around flat space was already considered in some detail in
Ref. \cite{hw93}, and shown to agree with the weak field expansion
in the continuum.
Here we provide a very short summary of the methods and results of this work.
Following Ref. \cite{rowi81}, one takes as background space a network
of unit cubes divided into tetrahedra by drawing in parallel sets of face
and body diagonals, as shown in Figure 8.
With this choice, there are $2^d-1=7$ edges per lattice point
emanating in the
positive lattice directions: three body principals, three
face diagonals and one
body diagonal, giving a total of seven components per lattice point.

\vskip 10pt

\begin{center}
\epsfxsize=7cm
\epsfbox{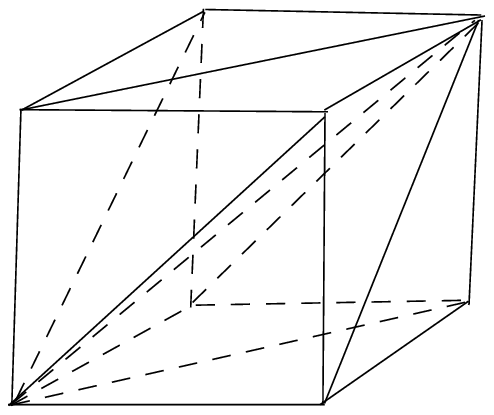}
\end{center}

\noindent{\small Figure 8. A cube divided into simplices.}

\vskip 10pt

It is convenient to use a binary notation for edges, so that the
edge index corresponds to the lattice direction of the edge, expressed as a
binary number
\bea
 (0,0,1) & \rightarrow & 1  \;\;\;\;  (0,1,1)  \rightarrow 3 \;\;\;\;   (1,1,1) \rightarrow 7 
\nonumber \\
 (0,1,0) & \rightarrow & 2  \;\;\;\;   (1,0,1)  \rightarrow 5
\nonumber \\
 (1,0,0) & \rightarrow & 4   \;\;\;\;  (1,1,0)  \rightarrow 6
\eea
The edge lengths are then allowed to fluctuate around their flat 
space values $l_i = l_i^0 (1 + \epsilon_i)$, and the second variation of the 
action is expressed as a quadratic form in $ \epsilon $
\beq
\delta^2 I = \sum_{m n} \; \epsilon^{(m)\; T} ~ M^{(m, n)} ~ \epsilon^{(n)} ,
\eeq
where $n,m$ label the sites on the lattice, and $M_{m n}$ is some Hermitian matrix.
The general aim is then to show that the above quadratic form is
equivalent to the expansion of the continuum Einstein-Hilbert action
to quadratic order in the metric fluctuations.
The infinite-dimensional matrix $M^{(m, n)}$ is best studied by going to momentum space;
one assumes that the fluctuation $\epsilon_i$ 
at the point $j$ steps from the origin in one coordinate direction,
$k$ steps in another coordinate direction, and $l$ steps in the third
coordinate direction, is related to the 
corresponding fluctuation $\epsilon_i$ at the origin by 
\beq
\epsilon_i^{(j+k+l)}= \omega_1^j ~ \omega_2^k \omega_4^l ~ \epsilon_i^{(0)} \; ,
\eeq
with $\omega_i=e^{i k_{i} }$.
In the smooth limit, $\omega_i=1+ i k_{i} +O(k_i^2)$, 
the lattice action and the continuum action are then expected to agree.
Note also that it is convenient here to set the lattice spacing in the three
principal directions equal to one; it can always be restored at the
end by using dimensional arguments.

It is desirable to express the lattice action in terms of
variables which are closer to the continuum ones, such as
$h_{\mu\nu}$ or 
$\bar h_{\mu\nu} = h_{\mu\nu} - {2 \over 3} \eta_{\mu\nu} h_{\lambda\lambda}$.
Up to terms that involve derivatives of the metric (and which reflect
the ambiguity of where precisely on the lattice the continuum metric
should be defined),
this relationship can be obtained by considering one tetrahedron,
and using the expression for the invariant line element
$ d s^2 = g_{\mu\nu} dx^{\mu} dx^{\nu} $
with $ g_{\mu\nu} = \eta_{\mu\nu} + h_{\mu\nu} $, where
$ \eta_{\mu\nu} $ is the diagonal flat metric.
Inserting $l_i \; = \; l_i^0 \; ( 1 + \epsilon_i )$,
with $l_i^0 = 1, \sqrt{2}, \sqrt{3} $ for the body principal
($i=1,2,4$), face diagonal ($i=3,5,6$), and body diagonal ($i=7$),
respectively, one obtains
\bea
( 1 + \epsilon_1 )^2 & \; = \; & 1 + h_{11}
\nonumber \\
( 1 + \epsilon_2 )^2 & \; = \; & 1 + h_{22}
\nonumber \\
( 1 + \epsilon_4 )^2 & \; = \; & 1 + h_{33}
\nonumber \\
( 1 + \epsilon_3 )^2 & \; = \; & 1 + {1 \over 2} ( h_{11} + h_{22} ) + h_{12}
\nonumber \\
( 1 + \epsilon_5 )^2 & \; = \; & 1 + {1 \over 2} ( h_{11} + h_{33} ) + h_{13}
\nonumber \\
( 1 + \epsilon_6 )^2 & \; = \; & 1 + {1 \over 2} ( h_{22} + h_{33} ) + h_{23}
\nonumber \\
( 1 + \epsilon_7 )^2 & \; = \; & 1 + {1 \over 3} ( h_{11} + h_{22} + h_{33} ) 
+ {2 \over 3} ( h_{12} + h_{23} + h_{13} ) 
\eea
(note that we use the binary notation for edges, but maintain the
usual index notation for the field $h_{\mu\nu}$).
The above relationship can then be inverted to give the $\epsilon$'s in term
of the $h$'s .
Note that there are seven $\epsilon_i$ variables, but only six 
$h_{\mu\nu}$'s (in general in $d$ dimensions we have
$2^d-1$ $\epsilon_i$ variables and $d(d+1)/2$ $h_{\mu\nu}$'s, which
leads to a number of redundant lattice variables for $d > 2$).

Thus, to lowest order in $h_{\mu\nu}$, one can perform a field rotation
on the lattice in order to go from the
$\epsilon_i$ variables to the $h_{\mu\nu}$'s (or $\bar h_{\mu\nu}$'s),
\beq
\epsilon^T \; M_{\omega} \; \epsilon \; = \; 
( \epsilon^T V^{\dagger -1} ) \; V^{\dagger} \;
M_{\omega} \; V \; ( V^{-1} \epsilon ) ,
\eeq
with
\beq
\epsilon = U_1 h \;\;\;\; h = U_2 \bar h \; ,
\eeq
and so
\beq
\epsilon \; = \;  V \bar h   \;\;\;\;  V = U_1 U_2    \; .
\eeq

Here $V$ and $U_1$ are $7 \times 6$ matrices, while $U_2$ is a 
$6 \times 6$ matrix,
\beq
U_1 = \left (
  \matrix{ {1\over 2} & 0 & 0 & 0 & 0 & 0 \cr 0 & 
   {1\over 2} & 0 & 0 & 0 & 0 \cr 0 & 0 & 
   {1\over 2} & 0 & 0 & 0 \cr {1\over 4} & {1\over 4} & 0 & 
   {1\over 2} & 0 & 0 \cr {1\over 4} & 0 & {1\over 4} & 0 & 
   {1\over 2} & 0 \cr 0 & {1\over 4} & {1\over 4} & 0 & 0 & 
   {1\over 2} \cr {1\over 6} & {1\over 6} & {1\over 6} & {1\over 3} & 
   {1\over 3} & {1\over 3} \cr  }  \right ) ,
\eeq
\beq
U_2 = \left (
  \matrix{{1\over 3} & -{2\over 3} & -{2\over 3} & 0 & 0 & 0 \cr -{2\over 3}
    & {1\over 3} & -{2\over 3} & 0 & 0 & 0 \cr -{2\over 3} & -{2\over 3} & 
   {1\over 3} & 0 & 0 & 0 \cr 0 & 0 & 0 & 
    1 & 0 & 0 \cr 0 & 0 & 0 & 0
    &  1 & 0 \cr 0 & 0 & 0 & 0 & 
   0 &  1 \cr  }  \right ) \; .
\eeq
The above rotation is an essential step in transforming the lattice action into
a form that looks like the continuum action, to quadratic order in the weak fields.
For the Regge-Einstein term the matrix $M_\omega$ describing the small fluctuations around
flat space is given by
\bea
( M_\omega )_{1,1} = & \; -2
\nonumber \\
( M_\omega )_{1,2} = & \; - \omega_1 \omega_4  - \bar \omega_2 \bar \omega_4
\nonumber \\
( M_\omega )_{1,4} = & \; 2 + 2 \bar \omega_2
\nonumber \\
( M_\omega )_{1,6} = & \; 2 \omega_1 + 2 \bar \omega_2 \bar \omega_4
\nonumber \\
( M_\omega )_{1,7} = & \; -3 \bar \omega_2 - 3 \bar \omega_4
\nonumber \\
( M_\omega )_{4,4} = & \; -8
\nonumber \\
( M_\omega )_{4,5} = & \; -4 \omega_2 - 4 \bar \omega_4
\nonumber \\
( M_\omega )_{4,7} = & \; 6 + 6 \bar \omega_4
\nonumber \\
( M_\omega )_{7,7} = & \; -18
\eea
with the remaining matrix elements obtained by permuting the appropriate indices.
Due to its structure, which is of the form
\beq
M_{\omega} = \left ( \matrix{ A_6 & b \cr b^{\dagger} & -18 \cr } \right ) ,
\eeq
where $A_6$ is a $6 \times 6$ matrix,
a rotation can be done which completely
decouples the fluctuations in $\epsilon_7$,
\beq
M^{\prime}_{\omega} = S^{\dagger}_{\omega} M_{\omega} S_{\omega} =
\left ( \matrix{ 
A_6 + {1 \over 18} b b^{\dagger} & 0 \cr 0 & -18 \cr } \right ) ,
\eeq
with 
\beq
S_{\omega}
= \left ( \matrix{ I_6 & 0 \cr {1 \over 18} b^{\dagger} & 1 \cr } \right ) \; .
\eeq
One then finds the first important result, namely that the small fluctuation matrix 
$M^{\prime}_{\omega}$ has three zero eigenvalues, corresponding to the 
translational zero modes for $M_{\omega}$
\beq
\left (  \matrix{ \epsilon_1 \cr \epsilon_2 \cr \epsilon_4 \cr
\epsilon_3 \cr \epsilon_5 \cr \epsilon_6 \cr \epsilon_7 \cr } \right ) =
\left (  \matrix{
1- \omega_1 & 0 & 0 \cr
0 & 1 - \omega_2 & 0 \cr
0 & 0 & 1 - \omega_4 \cr
{1 \over 2} ( 1 - \omega_1 \omega_2 ) & 
{1 \over 2} ( 1 - \omega_1 \omega_2 ) & 0 \cr
{1 \over 2} ( 1 - \omega_1 \omega_4 ) &  0 &
{1 \over 2} ( 1 - \omega_1 \omega_4 ) \cr
0 & {1 \over 2} ( 1 - \omega_2 \omega_4 ) & 
{1 \over 2} ( 1 - \omega_2 \omega_4 ) \cr 
{1 \over 3} ( 1 - \omega_1 \omega_2 \omega_4 ) & 
{1 \over 3} ( 1 - \omega_1 \omega_2 \omega_4 ) & 
{1 \over 3} ( 1 - \omega_1 \omega_2 \omega_4 ) \cr } \right ) 
\times 
\left (  \matrix{ x_1 \cr x_2 \cr x_3 \cr } \right ) \; 
\eeq
where $x_1, x_2, x_3$ are three arbitrary parameters.
The remaining eigenvalues are $-18$ (once) and $O(k^2)$ (three times).
Notice that one mode, corresponding to the fluctuations in the
body diagonal $\epsilon_7$, completely decouples.
The next step is to transform the lattice weak field action into a form similar 
(in fact identical) to the continuum form.
One further rotation by the ($6 \times 6$) matrix $T_\omega$,
defined by
\beq
T_\omega =  \left (
  \matrix{ {{{\omega_1}}\over 6} & -{{{\omega_1}}\over 3} & -{{{\omega_1}}\over 3}
    & 0 & 0 & 0 \cr -{{{\omega_2}}\over 3} & {{{\omega_2}}\over 6} & 
   -{{{\omega_2}}\over 3} & 0 & 0 & 0 \cr -{{{\omega_4}}\over 3} & 
   -{{{\omega_4}}\over 3} & {{{\omega_4}}\over 6} & 0 & 0 & 0 \cr 
   -{{  {\omega_1} {\omega_2}   }\over {12}} & 
   -{{  {\omega_1} {\omega_2}   }\over {12}} & 
   -{{  {\omega_1} {\omega_2}   }\over 3} & {1\over 2} & 0 & 0 \cr 
   -{{  {\omega_1} {\omega_4}   }\over {12}} & 
   -{{  {\omega_1} {\omega_4}   }\over 3} & 
   -{{  {\omega_1} {\omega_4}   }\over {12}} & 0 & {1\over 2} & 0 \cr 
   -{{  {\omega_2} {\omega_4}   }\over 3} & 
   -{{  {\omega_2} {\omega_4}   }\over {12}} & 
   -{{  {\omega_2} {\omega_4}   }\over {12}} & 0 & 0 & {1\over 2} \cr 
   0 & 0 & 0 & 0 & 0 & 0 \cr  }  \right )
\eeq
gives the new small fluctuation matrix
\beq
L_{\omega} \; = T_{\omega}^{\dagger} \; M^{\prime}_{\omega} \; T_{\omega} .
\eeq
This last transformation is equivalent to a change to trace-reversed
metric variables.
Finally one defines the gauge fixed matrix
\beq
\tilde L_{\omega} \; = \; L_{\omega} - { 1 \over 2 } C^{\dagger}_\omega C_\omega ,
\eeq
where $C_\omega$ is introduced in order to give the lattice analog of
the harmonic gauge fixing term, with
\beq
C_\omega = {1 \over 6 } \left (
  \matrix{ 5 \left( -1 + {\omega_1} \right)  & 1 - {\omega_1} & 1 - {\omega_1} & 
   6 \left( 1 - {1\over {{\omega_2}}} \right)  & 
   6 \left( 1 - {1\over {{\omega_4}}} \right)  & 0 \cr 1 - {\omega_2} & 
   5 \left( -1 + {\omega_2} \right)  & 1 - {\omega_2} & 
   6 \left( 1 - {1\over {{\omega_1}}} \right)  & 0 & 
   6 \left( 1 - {1\over {{\omega_4}}} \right)  \cr 1 - {\omega_4} & 1 - {\omega_4}
    & 5 \left( -1 + {\omega_4} \right)  & 0 & 
   6 \left( 1 - {1\over {{\omega_1}}} \right)  & 
   6 \left( 1 - {1\over {{\omega_2}}} \right)  \cr  } \right ) .
\eeq
Then the form of $\tilde L_{\omega}$ is precisely equivalent
to the corresponding continuum expression, in trace-reversed variables
and in the harmonic gauge \cite{hw93}.
The seemingly complex combined $S_{\omega}$ and $T_{\omega}$
rotations just correspond to a rotation, from the original lattice edge fluctuation
variables ($\epsilon$) to the trace reversed metric variables ($\bar h$).

Perhaps the most important aspect of the above proof of convergence of
the lattice curvature term, and more generally of the whole lattice Wheeler-DeWitt
equation,  towards the corresponding continuum expression
is in its relevance to the weak field limit, to a perturbative expansion in $G$,
and to a $WKB$ expansion of the wavefunction.
The latter are all issues that have already earned some consideration in the continuum
formulation \cite{dew67,leu64,kuc76,bar98}.
The present work suggests that most of
those continuum results will remain valid, as long as they are derived in
the context of the stated approximations.
In particular, there is no reason to expect the {\it lattice} semiclassical wavefunction
to have a different form (apart from the use of different variables, whose 
correspondence has been detailed in this section) compared
to the continuum one \cite{kuc76}.


\vskip 20pt
\section{Conclusions}

In this paper we have presented a lattice version of the Wheeler-DeWitt equation of
quantum gravity. 
The present 3+1 approach is based on the canonical, and therefore Lorentzian, formulation of 
quantum gravity, and should therefore be regarded as complementary to the 
four-dimensional Euclidean lattice version of the same theory discussed earlier in other papers.
The equations are explicit enough to allow a number of potentially useful practical calculations,
such as the strong coupling expansion, mean field theory, and  the variational method.
In the preceding sections we have outlined a number of specific calculations to 
illustrate the mechanics of the lattice theory, and the likely physical interpretation of the results.
Because of its reliance on different set of nonperturbative approximation methods, the 
formulation presented here should be useful in viewing the older Euclidean lattice results from a very
different perspective; in a number of instances we have been able to show the physical
similarities between the two types of results.

Nevertheless the phenomenal complexity of the original continuum theory, and of the Euclidean
lattice approach, with all its issues of, for example, perturbative nonrenormalizability in four dimensions, has not gone away;
it just got reformulated in a rather different language involving a Schr\"odinger -like equation, wavefunctionals,  operators and states.
The hope is that the present approach will allow the use of a different set of approximation
methods, and numerical algorithms, to explore what in some instances are largely known issues, 
but now from an entirely different perspective.
Among the problems one might want to consider we list: the description of invariant correlation
functions \cite{cor94,gid06},
the behavior of the fundamental gravitational correlation length $\xi$ as a function of the coupling $G$, 
the approach to the lattice continuum limit at $G_c$, estimates for the critical exponents in the
vicinity of the fixed point, and the large-scale behavior of the gravitational Wilson loop \cite{hw07}.
Note that what is sometimes referred to as the problem of time does not necessarily affect
the above issues, which in our opinion can be settled by looking  exclusively at certain 
types of invariant correlations along the spatial directions only.
These Green's functions should then provide adequate information about the nature
of correlations in the full theory, without ever having to make reference to a time variable.


\vspace{20pt}

{\bf Acknowledgements}

The authors wish to thank Thibault Damour and the
I.H.E.S. in Bures-sur-Yvette for a very warm hospitality. 
Discussions with Reiko Toriumi are gratefully acknowledged.
The work of HWH was supported in part by the I.H.E.S. and
the University of California.
The work of RMW was supported in 
part by the UK Science and Technology Facilities Council.

\newpage

\appendix

\section*{Appendix}

\vskip 20pt
\newsection{The matrix $\bf M$ in 3+1 dimensions}
\label{sec:matrix}
\hspace*{\parindent}

The matrix of coefficients of the second partial derivative operators
in 3+1 dimensions is given by $ {\bf M} \over {24 V}$, where $\bf M$ is
a symmetric $6 \times 6$ matrix with entries as follows:
\bea
M_{11}& = &2 a^2
\nonumber \\
M_{12}& = &a^2 + b^2 + c^2 - 2 a b - 2 b 
\nonumber \\
M_{13}& = &a^2 + d^2 + e^2 - 2 a e - 2 d e
\nonumber \\
M_{14}& = &a^2 + b^2 + c^2 - 2 a c - 2 b c
\nonumber \\
M_{15}& = &a^2 + d^2 + e^2 - 2 a d - 2 d e
\nonumber \\
M_{16}& = &b^2 + c^2 + d^2 + e^2 - 2 a f + 2 b d + 2 c e - 2 b c - 2 b e
- 2 c d - 2 d e 
\nonumber \\
M_{22}& = &2 c^2
\nonumber \\
M_{23}& = &c^2 + d^2 + f^2 - 2 c f - 2 d f
\nonumber \\
M_{24}& = &a^2 + b^2 + c^2 - 2 a b - 2 a c
\nonumber \\
M_{25}& = &a^2 + b^2 + d^2 + f^2 + 2 a f + 2 b d - 2 c e - 2 a b - 2 a d
- 2 b f - 2 d f
\nonumber \\
M_{26}& = &c^2 + d^2 + f^2 - 2 c d - 2 d f
\nonumber \\
M_{33}& = &2 d^2
\nonumber \\
M_{34}& = &a^2 + c^2 + e^ + f^2 + 2 a f - 2 b d + 2 c e - 2 a c - 2 a e
- 2 c f - 2 e f
\nonumber \\
M_{35}& = &a^2 + d^2 + e^2 - 2 a d - 2 a e
\nonumber \\
M_{36}& = &c^2 + d^2 + f^2 - 2 c d - 2 c f
\nonumber \\
M_{44}& = &2 b^2
\nonumber \\
M_{45}& = &b^2 + e^2 + f^2 - 2 b f - 2 e f
\nonumber \\
M_{46}& = &b^2 + e^2 + f^2 - 2 b e - 2 e f
\nonumber \\
M_{55}& = &2 e^2
\nonumber \\
M_{56}& = &b^2 + e^2 + f^2 - - 2 b e - 2 b f
\nonumber \\
M_{66}& = &2 f^2 \, .
\eea

\vskip 40pt


\vskip 20pt
\newsection{ Lattice Hamiltonian for Gauge Theories }
\hspace*{\parindent}
\label{sec:lgt}

It is of interest to see how the Hamiltonian approach has fared
for ordinary $SU(N)$ gauge theories, whose nontrivial infrared 
properties (confinement, chiral symmetry breaking) cannot be
seen to any order in perturbation theory, and require therefore
some sort of nonperturbative approach, based for example
on the strong coupling expansion.
In the continuum one starts from the Yang-Mills action
\beq
I \; = \; - { 1 \over 4 g^2 } \int d^4 x \, F_{\mu\nu}^a \, F^{\mu\nu a}
\label{eq:ym-action}
\eeq
with field strength
\beq
F_{\mu\nu}^a = \partial_\mu A_\nu^a - \partial_\nu A_\mu^a
+ g f_{abc} \, A_\mu^b A_\nu^c
\eeq
and gauge fields $A_\mu^a$  with $(a=1\dots N^2-1)$,
where the quantities $f_{abc}$ are the
structure constants of the Lie group, such that the generators
satisfy $[T_a,T_b]= i f_{abc} T_c$.

A lattice regularized form of the gauge action in 
Eq.~(\ref{eq:ym-action}) was given in Ref. \cite{wil74}, see also
\cite{bal75}.
The theory is defined on a $d$-dimensional hyper-cubic
lattice with lattice spacing $a$, 
vertices labeled by an index ${\bf n}$ and directions by ${\bf \mu}$.
The group elements $U_{n \mu} = \exp ( i a g \, A_\mu^a \, T_a )$ 
are defined in the fundamental representation, and reside
on the links of the lattice.
The pure gauge Euclidean action involves a sum of traces of 
path-ordered products 
[with $U_{- \mu} (n+ \nu )  = U^\dagger_\mu (n) $] 
of unitary $U_\mu (n)$ matrices around an elementary
square loop ("plaquettes", here denoted by $\Box$),
\beq
I [ U] \; = \; - { a^{4-d} \over 4 \, g^2 } \; \sum_{\Box} 
\tr \left [ U U U^\dagger U^\dagger \, + \, h.c. \right ] \; .
\label{eq:wilson-a}
\eeq
The action is locally gauge invariant with respect to the change
\beq
U_\mu (n) \; \rightarrow \; 
V^\dagger ( n ) \, U_{\mu} ( n ) \, V (n+ \nu) \; ,
\eeq
where $V$ is an arbitrary $SU(N)$ matrix defined on the lattice sites.

The next step is to define the path integral as
\beq
Z ( g^2 ) \; = \; \int [ d U_H ] \; \exp \left ( - I [U]  \right ) \; ,
\label{eq:wilson-z}
\eeq
where $[ d U_H ] $ is the Haar measure over the group $SU(N)$,
one copy for each lattice link variable $U$.
A lattice regularized Hamiltonian can then be defined on a purely spatial lattice, 
by taking the zero lattice spacing limit in the time direction \cite{ks75,kog83}.
Local gauge invariance further allows one to set all the link 
variables in the time direction to unity, $U_{n0}=1$, or 
$A_{n0}^a =0$ in this lattice version of the temporal gauge.
The $\dot{U}$ variables can now be eliminated
by introducing generators of local rotations
$E^a_i ( {\bf n} ) $, defined on the links (with spatial
directions labeled by $i,j=1,2,3$) and satisfying
the commutation relations
\beq
\left [  E^a_i ( {\bf n} ) , U_j ( {\bf m} ) \right ] \, = \, 
T^a \, U_i ( {\bf n} ) \, \delta_{ij} \, \delta_{\bf n m} \; ,
\label{eq:gen-u}
\eeq
along with the $SU(N)$ generator algebra commutation relation
\beq
\left [  E^a_i ( {\bf n}) , E^b_j ( {\bf m} ) \right ] \, = \, 
i \, f^{abc} \, E^c_i ( {\bf n} ) \, \delta_{ij} \, \delta_{\bf n m} \; .
\label{eq:lat-gen}
\eeq
This finally gives for the Hamiltonian of Wilson's lattice gauge theory \cite{ks75}.
\beq
H \; = \; { g^2  \over 2 a } \sum_{\rm links}  E^a E^a
\, - \, \sum_{\Box} { 1 \over 4 a g^2 } 
\tr \left [ U U U^\dagger U^\dagger \, + 
\, h.c. \right ]
\label{eq:ham-ks}
\eeq
The first term in Eq.~(\ref{eq:ham-ks}) is the lattice analog
of the electric field term ${\bf E}^2$, while the second term
is a lattice discretized version, involving lattice finite differences,
of the magnetic field $ ( \nabla \times {\bf A} )^2 $ term.
In this picture the analog of Gauss's law is a constraint which needs
to be enforced on physical states at each spatial site ${\bf n}$
\beq
\sum_{i=1}^6 E^a_i ( {\bf n} ) \, \vert \Psi \rangle \; = \;  0
\eeq
In general, and irrespective of the symmetry group chosen,
the ground state in the strong coupling $g^2 \rightarrow \infty$
limit has all the $SU(N)$  rotators in their ground state.
In this limit the Hamiltonian has the simple form
\beq
H_0 \; = \; { g^2  \over 2 a } \sum_{\rm links}  E^a_i E^a_i \; .
\label{eq:ham-strong}
\eeq
Then the vacuum is a state for which each link is in a color
singlet state
\beq
E^a_i \; \vert 0 \rangle  \; = \; 0 \; .
\eeq
The lowest order excitation of the vacuum is a state
 with one unit of chromo-electric field on each link
of an elementary lattice square, and energy 
\beq
E_\Box \; = \;  4 \cdot { g^2 \over 2 a } \, { N^2 - 1 \over 2 N }  \; .
\eeq
Raleigh-Schr\"odinger perturbation theory can then be
used to compute corrections to arbitrarily high order in $1/g^2$.
But ultimately one is interested in the limit $g^2 \rightarrow 0$,
corresponding to the ultraviolet asymptotic freedom fixed point of the 
non-Abelian gauge theory,
and thus to the lattice continuum limit $a \rightarrow 0$.
Thus in order to recover the original theory's continuum limit,
one needs to examine a limit where the mass gap in units
of the lattice spacing goes to zero, $ a \, m (g) \rightarrow 0$.
This limit then corresponds to an infinite correlation length in lattice
units; the zero lattice spacing limit so described is a crucial step
in fully recovering desirable properties (rotational or Lorentz invariance,
asymptotic freedom, massless perturbative gluon excitations etc.)
of the original continuum theory.

\vskip 40pt


\vfill


\end{document}